\documentclass[reprint,amsmath,amssymb]{revtex4-2}
\usepackage{xcolor}
\usepackage{graphicx}
\usepackage{physics}
\usepackage{comment}
\usepackage{todonotes}
\usepackage{enumitem}
\usepackage{hyperref}
\usepackage[sort]{cleveref}

\newcommand{\papertitle}{Universal scaling in the rheology of dense cellular systems}

\begin{document}

\title{\papertitle}
\author{Helen S. Ansell}
\affiliation{Department of Physics, Emory University, Atlanta, Georgia 30322}
\author{Daniel M. Sussman}
\email{daniel.m.sussman@emory.edu}
\affiliation{Department of Physics, Emory University, Atlanta, Georgia 30322}

\date{\today}

\begin{abstract}
Biological tissues must dynamically transition between rigid and fluid-like states during processes like morphogenesis and collective migration, often while simultaneously resisting physiological shear stresses.
It remains unclear whether these tissue dynamics are governed by the same non-equilibrium critical phenomena that control conventional disordered matter.
Here we show that model cell monolayers under constant stress display a rich phase diagram of nonlinear rheology.
In rigid regimes, small internal fluctuations maintain a solid-like state up to a finite yield stress, above which the tissue shear-thins; conversely, fluid-like regimes exhibit robust continuous and discontinuous shear thickening, culminating in structural arrest via shear jamming.
This space-filling shear-jamming transition is accompanied by structural changes including the formation of system-spanning force chains and the emergence of orientational ordering.
We demonstrate that the macroscopic viscosity across these disparate regimes is described by universal scaling behavior controlled by the same underlying physical parameters.
These results establish confluent tissues as a distinct class of disordered matter, demonstrating that universal jamming phenomena can emerge entirely through shape-driven topological constraints to regulate biological mechanics.

\end{abstract}

\maketitle

\section*{Introduction}

Epithelial monolayers---the simplest tissues in multicellular organisms---are essential for development and normal physiology.
To maintain biological function, they face a fundamental mechanical tension: they must withstand external physiological shear stresses (e.g., from blood flow or gut motility), yet remain capable of internal fluid flow during dynamic processes like collective cell migration and wound healing~\cite{Espina2023}.

Individual living cells behave as active soft glassy materials: increasing an initially low applied stress or strain causes the response to cross over from a linear viscoelastic regime to strongly nonlinear behavior (see, e.g., Ref.~\cite{Kollmannsberger2011Review} for a review).
Experimental challenges in probing the macroscopic rheology of cell collectives~\cite{karnat2025noninvasive,Harris2012,Mary2022,Duque2024,Elkins2015,Dakhil2016,Wu2018} have made it difficult to determine how single-cell mechanics dictate tissue-level flow.
Despite these challenges, dense collections of cells are known to exhibit a diverse array of complex viscoelastic and nonlinear rheological behaviors, including strain stiffening, yielding, and shear thinning~\cite{karnat2025noninvasive, mongera2018fluid,Duque2024,Fernandez2007,Tlili2020,Blanch2017,Harris2012,kiran2021effect,Sanematsu2021,Elkins2015,Dakhil2016,Mary2022,Wu2018,Islam2025}.
To understand tissue-scale rheology, highly coarse-grained cell models have uncovered many 
linear~\cite{Tong2023,Bonfanti2020,Nestor2018,Hernandez2022,Yang2017,Ishihara2017,Popovic2017,Anand2026} and nonlinear~\cite{Pasupalak2021,Huang2022,Fielding2023,Hertaeg2024,Duclut2021,Matoz-Fernandez2017,Popovic2021,Merzouki2016,Anand2026,Ghosh2026} features across different parameter regimes, including yielding and shear thinning~\cite{Matoz-Fernandez2017,Duclut2021,Ishihara2017,Hertaeg2024,Sanematsu2021,Popovic2021}, shear thickening~\cite{Hertaeg2024,Ghosh2026}, strain stiffening~\cite{Ameen2026}, and shear-induced solidification~\cite{Huang2022}.

Complex nonlinear rheology is a hallmark of disordered materials, including granular matter and dense suspensions, driven by competing effects such as packing density, inter-particle interactions and imposed shear.
In equilibrium systems, scaling analyses near critical points rigorously identify the physical principles governing phase behavior.
Extending these approaches to non-equilibrium steady states has revealed that the rheology of sheared granular matter is consistent with critical scaling near a jamming transition~\cite{Olsson2007, vaagberg2014dissipation}.
More recently, similar scaling frameworks have successfully described shear thickening and shear jamming in dense suspensions~\cite{Ramaswamy2023,Malbranche2022Frontiers,Ramaswamy2025,Barth2026,Mao2026,Bhowmik2026}.
Because cell monolayers are space-filling, lacking interstitial fluid and inter-particle friction, it is unknown if they obey the same universal scaling as these particulate systems.
If they do, how does the seemingly similar interplay between an underlying fluidity transition, internal fluctuations, and external stresses give rise to such diverse viscoelastic flow responses?

Here, we investigate the steady-state flow response of cell monolayers under constant shear stresses using simulations of the 2D Voronoi cell model.
In this highly coarse-grained system, the non-equilibrium motion of individual cells can be approximated as random internal fluctuations that act as an effective temperature $T$~\cite{fletcher2014vertex, alert2020physical}, while the complex competition between biological processes like cell-cell adhesion and active cortical tensions is encoded in a preferred cell-shape parameter $p_0$~\cite{Farhadifar2007,Staple2010,Bi2015,Bi2016}. 
By varying $p_0$ and $T$, we span a wide range of underlying tissue states.

In regimes where the cell shape dictates a mechanically rigid structure, the tissue exhibits a solid-like state with finite yield stress when the internal fluctuations are sufficiently small; 
above the yield stress, external driving induces shear thinning.
By contrast, the regime where the underlying system is fluid-like shows continuous and discontinuous shear thickening, culminating in structural arrest via \emph{shear jamming}.
This observation is notable because the typical dense-suspension picture of shear jamming, which is understood to result from frictional contacts overcoming lubrication forces~\cite{Wyart2014, Brown2014}, cannot apply in these space-filling model cellular systems.

We demonstrate that a universal scaling form collapses the macroscopic viscosity in these seemingly disparate regimes, highlighting that the same underlying physical parameters control the observed nonlinear rheology.
Finally, we identify structural changes as the system shear jams, including the formation of system-spanning force chains~\cite{Nampoothiri2020,Vinutha2023,Countryman2025} and the emergence of cell elongation and orientational alignment.
Together, our results establish universal features and structural signatures of model cell monolayers that provide a framework for interpreting the rheology of experimental cell systems.

\section*{Results}

\subsection*{Nonlinear rheology}

\begin{figure*}
\centering
\includegraphics[width=\textwidth]{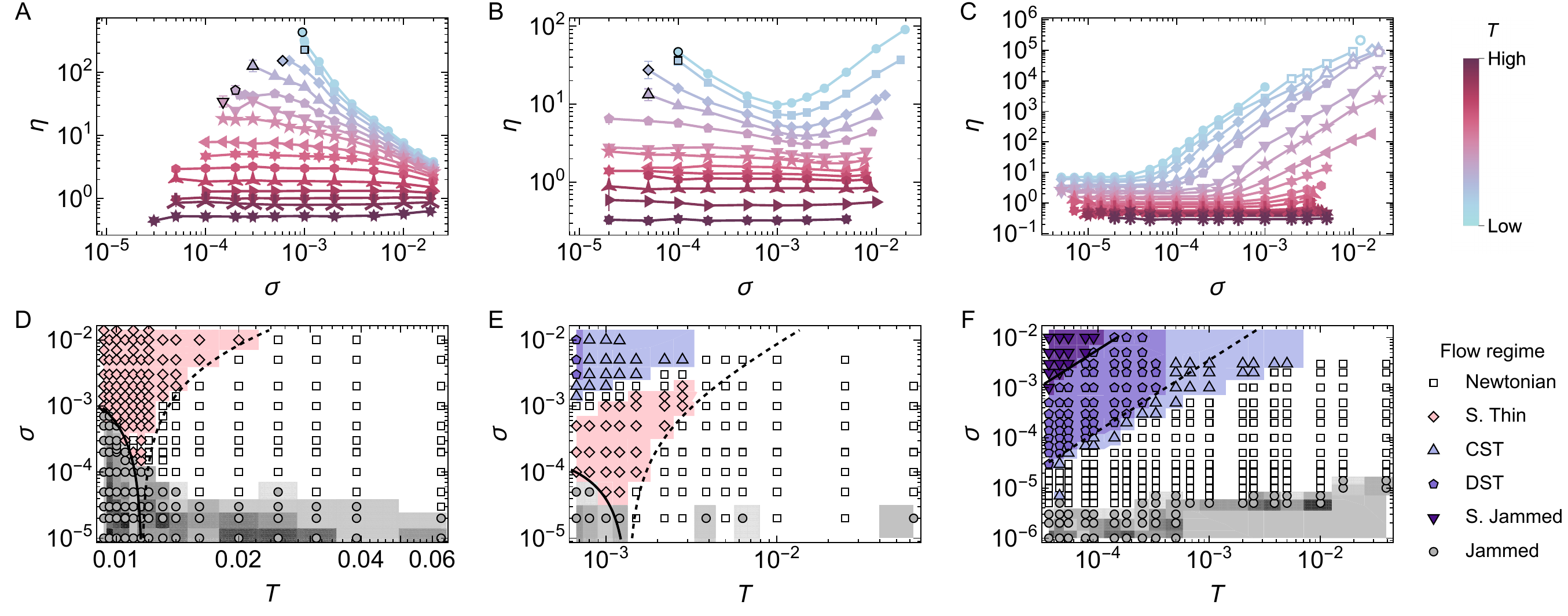}
\caption{
    \textbf{Steady-state flows of the Voronoi cell model reveal distinct regimes of yielding, shear thinning, shear thickening, and shear jamming.}
    (A--C) Measured viscosity $\eta$ for different imposed shear stresses $\sigma$ and temperatures $T$ for (A) $p_0=3.75$, (low $p_0$), (B) $p_0=3.815$ (intermediate $p_0$), and (C) $p_0=3.85$ (high $p_0$).
    Curve colors indicate temperatures (see scale bar), which correspond to $1\lesssim\tau_{\alpha}\lesssim 10^4$ for a given $p_0$, where $\tau_{\alpha}$ is the characteristic relaxation time.
    Black outlined markers indicate the approximate yield stress $\sigma_y$ in the low-$T$ regime. Points sampled with $\sigma<\sigma_y$ did not produce a flow.
    Markers with white outlines in C indicate points with $\eta>\eta^*$, which we identify as shear jammed.
    Note that data in this shear-jammed regime for which no flow is observed is not plotted.
    (D--F) Corresponding flow phase diagrams for each $p_0$, highlighting the different rheological regimes. 
    Solid and dashed black lines in D--F respectively indicate the predicted boundaries for the yielding and nonlinear flow transitions (\cref{eq:yield-boundary,eq:nonnewt}).
    Background shading for gray points indicates the proportion of samples at that state point that gave a flow, with lighter shading indicating a higher proportion.
    }
\label{fig:flow-phase}
\end{figure*}

We use a standard implementation of the 2D Voronoi cell model of a system of $N=4096$ polygonal cells, each of which has a target area $A_0$ and a target perimeter $P_0$.
Deviations from these target parameters have an energetic cost $E = \sum_{i=1}^N E_i$. 
Here
\begin{equation}
E_i  = K_A\left(A_i-A_{0}\right)^2 + K_P\left(P_i-P_{0}\right)^2,
\end{equation}
where $A_i$ and $P_i$ are  the area and perimeter of cell $i$, respectively, and $K_A$ and $K_P$ are the corresponding elastic moduli.
Underlying features of this system are controlled by the characteristic cell shape index $p_0 = P_0/\sqrt{A_0}$.
Below a critical value $p_0^*\approx3.81$, at zero temperature the cells are geometrically frustrated from achieving their target shapes, leading to mechanical rigidity.
Conversely, increasing $p_0$ past $p_0^*$ progressively relieves this frustration; 
whereas some cell models have a precise mechanical transition at $p_0^*$~\cite{Bi2015}, the 2D Voronoi model has a broadened transition~\cite{Sussman2018SM}, and we find that a finite yield stress persists for $p_0\lesssim3.825$. 
Here we consider $p_0$ values spanning this transition, in the range $3.75\leq p_0\leq 3.85$. 

Biological tissues are inherently active systems, driven by processes like cell motility.
To isolate the fundamental physics of shape-driven rheology, we study 
this system in thermal equilibrium, where an explicit temperature quantifies the magnitude of these internal random fluctuations.
We first characterize the flow response by measuring the viscosity $\eta$ as the cell shape index $p_0$, temperature $T$ and shear stress $\sigma$ are varied.
For each $p_0$, the range of $T$ studied corresponds to characteristic alpha-relaxation times $\tau_{\alpha}$ of the unsheared state in the range $1\lesssim\tau_{\alpha}\lesssim 10^4$~\cite{Ansell2025}, which samples both solid-like and fluid-like behavior in the $p_0\lesssim p_0^*$ regime~\cite{Li2018}.
Details of the simulations are given in Methods.

We observe a wide range of nonlinear rheological phenomena as a function of $p_0$ and $T$, as shown in \cref{fig:flow-phase,fig:supp-flow-phase}.
At low $p_0$ ($p_0<p_0^*$), where the underlying system is mechanically rigid, the low-$T$ regime exhibits a finite yield stress $\sigma_y$ below which there is no flow. 
For $\sigma>\sigma_y$, solid-like behavior gives way to pronounced shear thinning, reminiscent of the Herschel-Bulkley rheology observed near the athermal jamming transition in granular materials~\cite{olsson2012herschel}. 
As $T$ increases, $\sigma_y$ decreases and the shear thinning becomes less pronounced, ultimately leading to a low-viscosity Newtonian flow regime at higher $T$.

The high-$p_0$ regime exhibits very different flow.
For $p_0=3.85$, where the underlying system is fluid-like, Newtonian flow is observed in the low-$T$, low-$\sigma$ regime.
This transitions to pronounced shear thickening with increasing $\sigma$. 
The flow curves (\cref{fig:supp-gamma-sigma}) show an an initial continuous shear thickening (CST) that becomes discontinuous shear thickening (DST), marked by a discontinuity in the shear rate $\dot\gamma$.
At high enough $\sigma$, $\dot\gamma$ decreases until the flow ceases entirely, indicating shear jamming.
In our data, this is characterized by $\dot\gamma$ being unmeasurable, meaning $\eta$ is undefined.
Because our steady-state protocol can sometimes give a large, finite $\eta$ even in solid-like regimes (similar to experimental systems~\cite{Peters2016,Dhar2020}), we introduce an empirical threshold $\eta^*=8000$ to define the shear-jammed state, alongside the high-$\sigma$ points where no flow is observed.
We validate this threshold by confirming that nearby regions of phase space also yield undefined $\eta$ and vanishingly small $\dot{\gamma}$, consistent with a shear-jammed state.

For intermediate $p_0$ values with $p_0>p_0^*$, the low-$T$ rheology shows a crossover between these different flow regimes.
Finite $\sigma_y$ and shear thinning are observed for $p_0\leq 3.825$, becoming weaker with increasing $p_0$, while shear thickening and shear jamming become increasingly pronounced with increasing $p_0$.
At high $T$, the flow becomes Newtonian across all $p_0$ values. 
The resulting flow phase diagrams for each $p_0$ are shown in \cref{fig:flow-phase,fig:supp-flow-phase}; details of the classification of points are given in Methods.

Interpreting these distinct flow regimes requires accounting for strong finite-size effects (see \cref{fig:supp-fss}).
In the low-$p_0$ regime, larger systems exhibit systematically lower yield stresses; at high $p_0$, small systems are prone to spurious structural arrest.
These finite-size effects become negligible at high $T$, but because small $N$ can artificially exaggerate rigidity, we employ $N=4096$ cells to ensure our observations capture genuine thermodynamic phase behavior.
Finally, across all $p_0$, we observe a low-stress resolution limit inherent to our protocol~\cite{Kuang2012}, which we distinguish from a true, thermodynamic yield stress.

\subsection*{Characterizing the low-$\sigma$ behavior}

\begin{figure}[h]
\centering
\includegraphics[width=\linewidth]{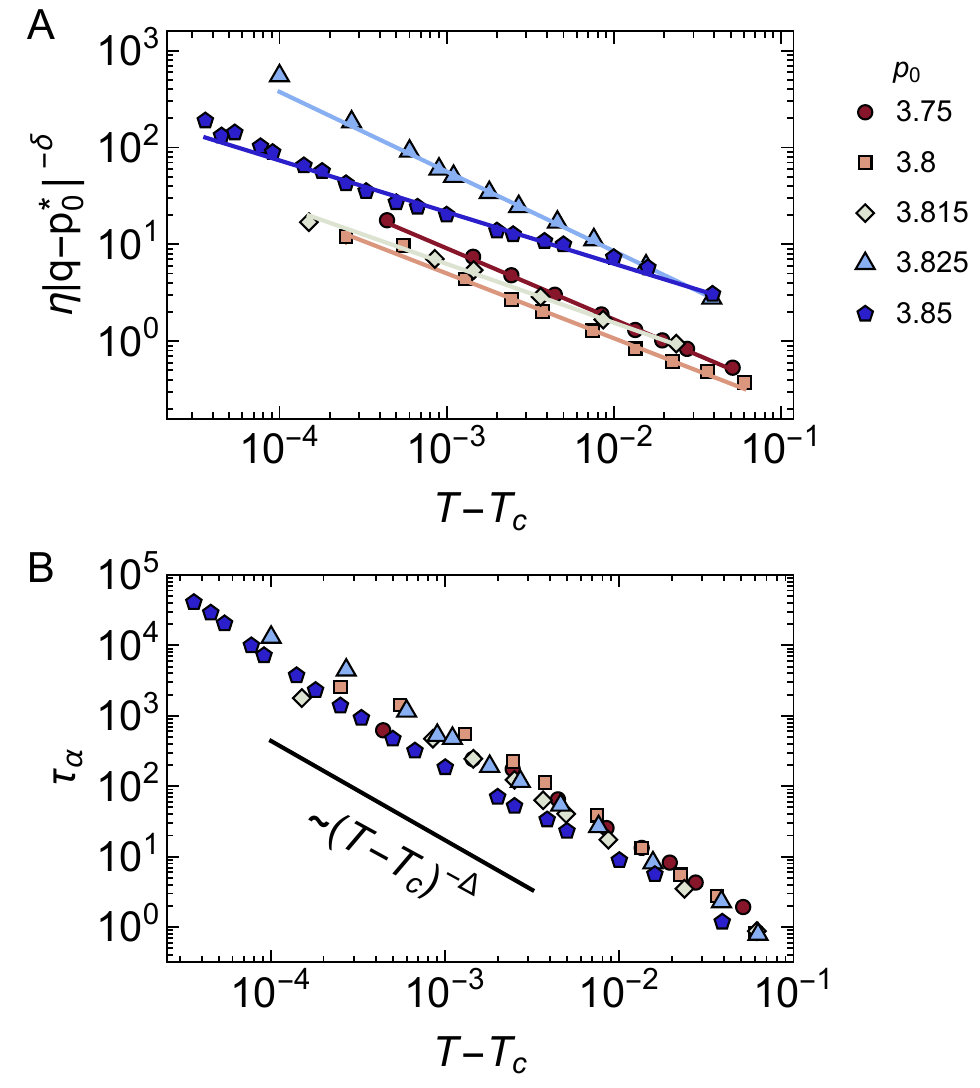}
\caption{
    \textbf{The low-stress rheology and intrinsic relaxation time exhibit power-law-like scaling near the fluidity transition.}
    (A) Dependence of $\eta \abs{q-p_0^*}^{-\delta}$ on $T-T_c$ in the low-$\sigma$ limit. Best fit lines are used to determine the exponent $\beta$. Here $\delta=0$ for $p_0<p_0^*$, while $\delta>0$ for $p_0>p_0^*$.
    (B) Scaling of the $\alpha$-relaxation time $\tau_{\alpha}$ with $T-T_c$ for $T>T_c$. The black line shows the mean slope, which defines the exponent $\Delta$. Power-law fits for each $p_0$ are shown in \cref{fig:supp-tau-alpha}.
}
\label{fig:scaling-low}
\end{figure}

The above results reveal a rich, multiphase rheology across different applied stresses, cell shape parameters, and temperatures.
To build a unified physical framework capable of capturing these disparate regimes, we systematically analyze the asymptotic limits of the parameter space. 
We begin by focusing on the low-$\sigma$ regime, in which the system transitions between a low-$T$ solid state and a higher-$T$ flowing state for $p_0\leq 3.825$ ($p_0=3.85$ flows across the full range of $T$). 
This motivates the ansatz 
\begin{equation}
\eta\sim
\begin{cases} 
(T-T_c)^{-\beta}&\text{if }T>T_c,\\
\infty &\text{otherwise,}
\end{cases}
\label{eq:eta-low-sigma}
\end{equation}
where the critical temperature $T_c$ bounds the regime exhibiting a finite yield stress.
We define $T_c$ by inspecting the flow curves (\cref{fig:flow-phase,fig:supp-flow-phase,fig:supp-gamma-sigma}): we identify the crossover from a finite-$\sigma$ intercept ($\sigma_y>0$) to a flowing state with $\dot{\gamma}\to 0$ as $\sigma \to 0$, and further refine the estimate by optimizing the fit to \cref{eq:eta-low-sigma}.
As shown in \cref{fig:scaling-low}A, this ansatz successfully describes the data for $p_0<p_0^*$, with the slope of the fit line giving the exponent $\beta$.

For $p_0>p_0^*$, the ansatz in \cref{eq:eta-low-sigma} instead gives two distinct regimes as $T-T_c$ varies (see \cref{fig:supp-p385-coll-no-q}A).
We find that the low-$\sigma$ viscosity scaling is better described by modifying the ansatz in \cref{eq:eta-low-sigma} to
\begin{equation}
\eta\sim(T-T_c)^{-\beta}\abs{q-p_0^*}^{\delta}
\label{eq:eta-low-sigma-q}
\end{equation}
for $T>T_c$, where $q = \expval{P_i/\sqrt{A_i}}$ is the mean cell shape index.
The quantity $q-p_0^*$, which depends on $p_0$, $T$, and weakly on $\sigma$ (see \cref{fig:supp-q}), has previously been shown to play an important role in the rheology of the Voronoi model under athermal quasistatic shear~\cite{Huang2022}.
For these $p_0$ values, we tune $\delta$ so that $\eta$ exhibits a single power-law regime, then perform a fit to determine $\beta$ (see \cref{fig:scaling-low}A).

The values of $T_c$, $\beta$ and $\delta$ are listed in \cref{tab:exponents} for each $p_0$.
Notably, these dynamical $T_c$ values are consistent with the emergence of orientational and translational ordering in the \emph{unsheared} system (see \cref{fig:supp-sf,fig:supp-order-param} and Methods), directly linking the onset of nonlinear flow to the underlying equilibrium phase behavior.
Furthermore, for $p_0=3.85$, the measured $T_c=0$ aligns with prior athermal studies showing that this system remains weakly jammed at zero temperature even above $p_0^*$~\cite{Sussman2018SM}.

The observation of power-law scaling of $\eta$ at low $\sigma$ for each $p_0$ is itself striking.
Most disordered materials display glassy dynamics characterized by an exponential (or super-exponential) dependence of $\eta$ on inverse temperature~\cite{Angell1995}, reflecting thermally activated relaxation across characteristic energy barriers.
In contrast, a power-law suggests a different, scale-free mechanism that lacks a characteristic energy scale.
Our observations may thus connect to emerging evidence of unusual sub-diffusive creep and fluid-like relaxation in unsheared dense tissues~\cite{shen2026dynamic}, and to findings of anomalous glassy dynamics in computational tissue models~\cite{Sussman2018EPL,Li2021,li2025connecting,Ansell2025}.
Previous studies have demonstrated power-law scaling of $\eta$ exclusively in the high-$p_0$ regime~\cite{Ansell2025}.
The persistence of this scaling at lower $p_0$ upon the introduction of a finite $T_c$ is surprising as it suggests a common underlying physical description governs the steady-state flow on both sides of the rigidity transition.

\subsection*{Universal scaling of rheological features}

Increasing $\sigma$ drives the system further from equilibrium, introducing a competition between the intrinsic relaxation dynamics and externally imposed driving.
This can be characterized by the product $\sigma\tau_{\alpha}$ (cf. the Weissenberg number in $\dot\gamma$-controlled rheology).
Because this quantity has dimensions of viscosity, non-dimensionalizing it by a characteristic zero-shear viscosity $\eta_0$ recovers a dimensionless stress ($\sigma \tau_\alpha / \eta_0 \sim \sigma / G_0$). 
Thus, $\sigma\tau_{\alpha}$ is the natural scaling variable governing departure from the low-$\sigma$ regime.

Examining the $T$ dependence of $\tau_{\alpha}$, we find that it scales algebraically, $\tau_{\alpha}\sim (T-T_c)^{-\Delta}$, with $\Delta\sim 1.4$ across all $p_0$ values (\cref{fig:scaling-low}B; see \cref{fig:supp-tau-alpha} for individual fits).
This power-law dependence is notable as it goes beyond the generic sub-Arrhenius scaling observed in this parameter regime~\cite{Sussman2018EPL,Li2021,Ansell2025,li2025connecting}, instead highlighting that the scaling of $\tau_{\alpha}$ could indicate slowing down near a critical temperature. 

Combining these identifies $\sigma\abs{T-T_c}^{-\Delta}$ as the  scaling variable governing departure from the low-$\sigma$ regime.
Building on our low-$\sigma$ scaling form (\cref{eq:eta-low-sigma-q}), we propose that the viscosity across the full range of $\sigma$ can be described by the scaling ansatz
\begin{equation}
\eta\sim\abs{T-T_c}^{-\beta}\abs{q-p_0^*}^{\delta}\mathcal{F}\left(\frac{\sigma}{\abs{T-T_c}^{\Delta}}\right),
\label{eq:scaling}
\end{equation}
where $\mathcal{F}$ is a scaling function.
For $p_0<p_0^*$, we have $\delta=0$, while $p_0>p_0^*$ has finite $\delta$.

\begin{figure}[t]
\centering
\includegraphics[width=\linewidth]{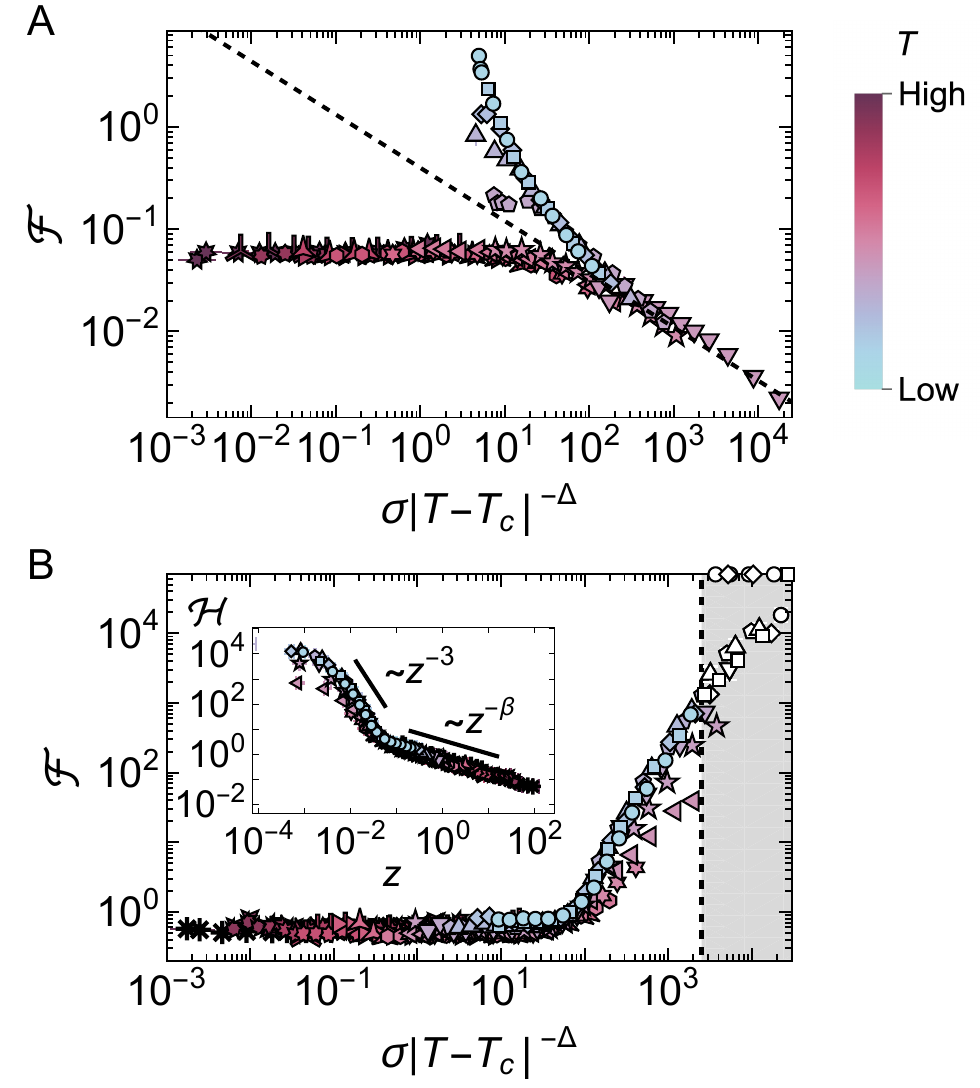}
\caption{
    \textbf{The diverse nonlinear flow regimes collapse onto universal scaling curves governed by the competition between external driving and structural relaxation.}
    (A) Collapse of the $p_0=3.75$ data, attained by plotting the scaling function $\mathcal{F}\sim\eta\abs{T-T_c}^{\beta}\abs{q-p_0^*}^{-\delta}$ as a function of the scaling variable $x = \sigma\abs{T-T_c}^{-\Delta}$ (\cref{eq:scaling}). 
    The dashed black line indicates the expected scaling of $x$ as $x\to\infty$.
    (B) Collapse of the $p_0=3.85$ data using the same scaling ansatz. 
    The vertical dashed line marks the critical value $x=x_c$ at structural arrest.
    White points are in the shear-jammed regime and have $\eta>\eta^*$. Those plotted on the horizontal axis indicate investigated stresses that yielded no measurable flow. 
    (inset) Cardy scaling near $x_c$ yields a scaling function $\mathcal{H}(z)$, with $z = \abs{x_c^{-1/\Delta}-x^{-1/\Delta}}$, diverging as $z^{-3}$.
    Values of $T_c$, $\delta$, $\beta$, and $\Delta$ are listed in \cref{tab:exponents}.
}
\label{fig:scaling}
\end{figure}

To test the validity of this ansatz, we first consider the low-$p_0$ regime, where the flow is characterized by yielding and shear thinning.
By plotting the scaled viscosity $\mathcal{F}(x) \sim \eta \abs{T-T_c}^{\beta}\abs{q-p_0^*}^{-\delta}$ against the scaling variable $x = \sigma\abs{T-T_c}^{-\Delta}$ for the $p_0=3.75$ data (\cref{fig:flow-phase}A), we observe an excellent data collapse (\cref{fig:scaling}A).
This collapse exhibits two distinct branches, characteristic of scaling on either side of a critical point.
Data points in the upper branch ($T<T_c$) that deviate slightly from the main collapse correspond to points very close to the yielding transition where the true viscosity may be higher than our estimates.
We note that the quality of the collapse is much more sensitive to the values of $T_c$ and $\beta$ than to the value of $\Delta$, within a reasonable range.

In the limit  $x\to\infty$, the two branches of $\mathcal{F}(x)$ approach a common curve that scales as $\mathcal{F}(x\to\infty)\sim x^{-\beta/\Delta}$, as indicated by the black dashed line in \cref{fig:scaling}A. 
This demonstrates that for $T=T_c$, $\eta(T=T_c)\sim\sigma^{-\beta/\Delta}$ as $\sigma\to 0$, as would be expected at a critical point.
The crossover exponent $\Delta > 0$ suggests that $\sigma$ is a relevant variable in the renormalization group sense. 
Consequently, the finite-$\sigma$ critical behavior should belong to a different universality class than the $\sigma\to 0$ limit.

The nature of the yielding transition at finite $\sigma$ depends on the behavior of the upper branch of $\mathcal{F}$, which diverges at a finite critical value $x_c\approx 4.9$. 
This implies that $\eta$ diverges whenever the scaling variable reaches $x_c$, defining a yield stress $\sigma_y$ that satisfies $\sigma_y\left(T_c-T\right)^{-\Delta} = x_c$.
This establishes a continuous line of transition points that predicts the yielding boundary:
\begin{equation}
\sigma_y(T) = x_c (T_c-T)^{\Delta}.
\label{eq:yield-boundary}
\end{equation}
For $\sigma\gtrsim10^{-4}$ (where a finite $\sigma_y$ is clearly resolved in our data), the prediction in \cref{eq:yield-boundary} (black curve, \cref{fig:flow-phase}D) agrees well with our numerical results.
These observations demonstrate that the scaling behavior for low $p_0$ is controlled by a finite-$T$ critical point at low $\sigma$ and a crossover to a different scaling regime at larger $\sigma$, in direct analogy with the density-dependent critical behavior near the jamming point of granular materials~\cite{Olsson2007}.

Turning now to the high-$p_0$ regime, we find that the same scaling ansatz (\cref{eq:scaling}) also successfully collapses the data in \cref{fig:flow-phase}C (see \cref{fig:scaling}B). 
The collapse exhibits a clear crossover:
In the small-$x$ Newtonian flow regime, $\mathcal{F}(x)$ is constant, demonstrating that the system obeys $\eta\sim \abs{T-T_c}^{-\beta}\abs{q-p_0^*}^{\delta}$.
Conversely, in the large-$x$ regime, $\mathcal{F}(x)$ increases dramatically, corresponding to the onset of shear thickening.
As for low-$p_0$, the fact that $\Delta>0$ means that $\sigma$ remains a relevant variable in the framework of the renormalization group.

To characterize the scaling as the system approaches the shear jammed state ($x\to x_c$), we first consider the divergence of $\mathcal{F}$.
Using our earlier criteria for identifying shear-jammed points, we find that $\mathcal{F}$ diverges at $x_c\sim 2500$ for $p_0=3.85$ (dashed line,  \cref{fig:scaling}B).
Since $T_c=0$ for $p_0=3.85$, we expect the system to shear jam along a line of points obeying $T_c^{\mathrm{SJ}}(\sigma) = (\sigma/x_c)^{1/\Delta}$, which is in good agreement with our numerical results (solid line, \cref{fig:flow-phase}F).

As $x$ approaches $x_c$, the scaling function $\mathcal{F}$ diverges as
\begin{equation}
\mathcal{F}(x)\sim (x_c-x)^{-\zeta}.
\end{equation}
Since $\mathcal{F}\sim\eta\abs{T-T_c}^{\beta}\abs{q-p_0^*}^{-\delta}$, this means that $\eta\sim(x_c-x)^{-\zeta}$ as $x\to x_c$.
To characterize this change in the viscosity scaling exponent, we follow the approach used in Refs.~\cite{Ramaswamy2023,Malbranche2022Frontiers,Ramaswamy2025,Barth2026} for dense suspensions and recast our scaling relation into the ``Cardy scaling form''~\cite{Cardy} 
\begin{equation}
\eta \sim \abs{q-p_0^*}^{\delta} \sigma^{-\beta/\Delta} \mathcal{H}\left(z\right)
\end{equation}
where $\mathcal{H}$ is a universal scaling function and $z=\abs{x^{-1/\Delta} - x_c^{-1/\Delta}}$.
We can therefore examine the scaling close to $x_c$ by plotting $\mathcal{H}(z)\sim\eta\abs{q-p_0^*}^{-\delta}\sigma^{\beta/\Delta}$ 
(see inset, \cref{fig:scaling}B).
Close to the shear-jamming transition ($z\to 0)$, we find $\mathcal{H}(z)\sim z^{-\zeta}$ with $\zeta\sim 3$. 
This is in stark contrast to larger $z$ ($x\ll x_c$), where the scaling is controlled by the $\sigma\to 0$ jamming point and $\mathcal{H}(z)\sim z^{-\beta}$ with $\beta\approx0.53$.
The change in $\zeta$ on the approach to shear jamming indicates that the divergence of $\eta$ in the sheared and unsheared regimes belongs to distinct universality classes.

Intermediate $p_0$ values show shear thinning and shear thickening at low $T$ as $\sigma$ is varied (see \cref{fig:flow-phase}B, \cref{fig:supp-flow-phase}).
We investigate the scaling of the shear-thinning and shear-thickening regimes separately for each $p_0$, including the Newtonian flow in both cases.
We find that \cref{eq:scaling} yields a good collapse in each regime (see \cref{fig:supp-scaling}), with qualitatively similar behavior to the collapses in \cref{fig:flow-phase}.
Interestingly, we find identical exponents ($\beta$, $\delta$, $\Delta$) for a given $p_0$, but the shear-thinning regime requires a higher $T_c$ than the shear-thickening regime (\cref{tab:exponents}).
For $p_0>p_0^*$ ($\delta\neq 0$), different combinations of $\beta$ and $\delta$ yield collapses of similar quality. 
The values presented here represent the estimates that optimize the collapse across both regimes.

In addition to predicting the boundaries between flowing and arrested states, the scaling collapse also predicts the boundary between Newtonian and non-Newtonian flow. 
Defining $x_n$ as the $x$ value at which $\mathcal{F}(x)$ starts to deviate from its low-$x$ constant plateau, we expect this boundary to be given by 
\begin{equation}\label{eq:nonnewt}
x_n = \sigma(T-T_c)^{-\Delta}.
\end{equation}
The resulting curves (dashed lines, \cref{fig:flow-phase,fig:supp-flow-phase}) show good agreement with our numerical phase boundaries.

\begin{table}[]
    \centering
    \begin{tabular}{c c c c c c}
         \hline\hline
          $p_0$ & $T_c^{\mathrm{Thin}}$ & $T_c^{\mathrm{Thick}}$ & $\beta$ & $\delta$ \\
         \hline
         3.75 & 0.0116 &  --   & 0.73 & 0 \\
         3.80 & 0.0026  & -- & 0.67  & 0\\
         3.815 & 0.0014 &   0.0004 & 0.61  & 0.2\\
         3.825 &  0.0004 &   0.00007  & 0.82  & 0.8\\
         3.85 & -- & 0 & 0.53  & 1.0\\
          \hline \hline
    \end{tabular}
    \caption{Summary of exponents used to give the best scaling collapse of the data for each $p_0$ using the scaling ansatz in \cref{eq:scaling}. $T_c^{\mathrm{Thin}}$ and $T_c^{\mathrm{Thick}}$ are respectively the $T_c$ values for the shear thinning and shear-thickening dominated regimes. We use $\Delta=1.4$ for all $p_0$ values.
    }
    \label{tab:exponents}
\end{table}

\subsection*{Structural signatures of shear jamming}
Having established a macroscopic scaling framework (\cref{eq:scaling}), we now characterize the corresponding cellular-level features.
We highlight distinct structural changes as the system shear thickens and undergoes structural arrest, establishing the shear-jammed regime as a distinct macroscopic state.

\begin{figure}
\centering
\includegraphics[width=\linewidth]{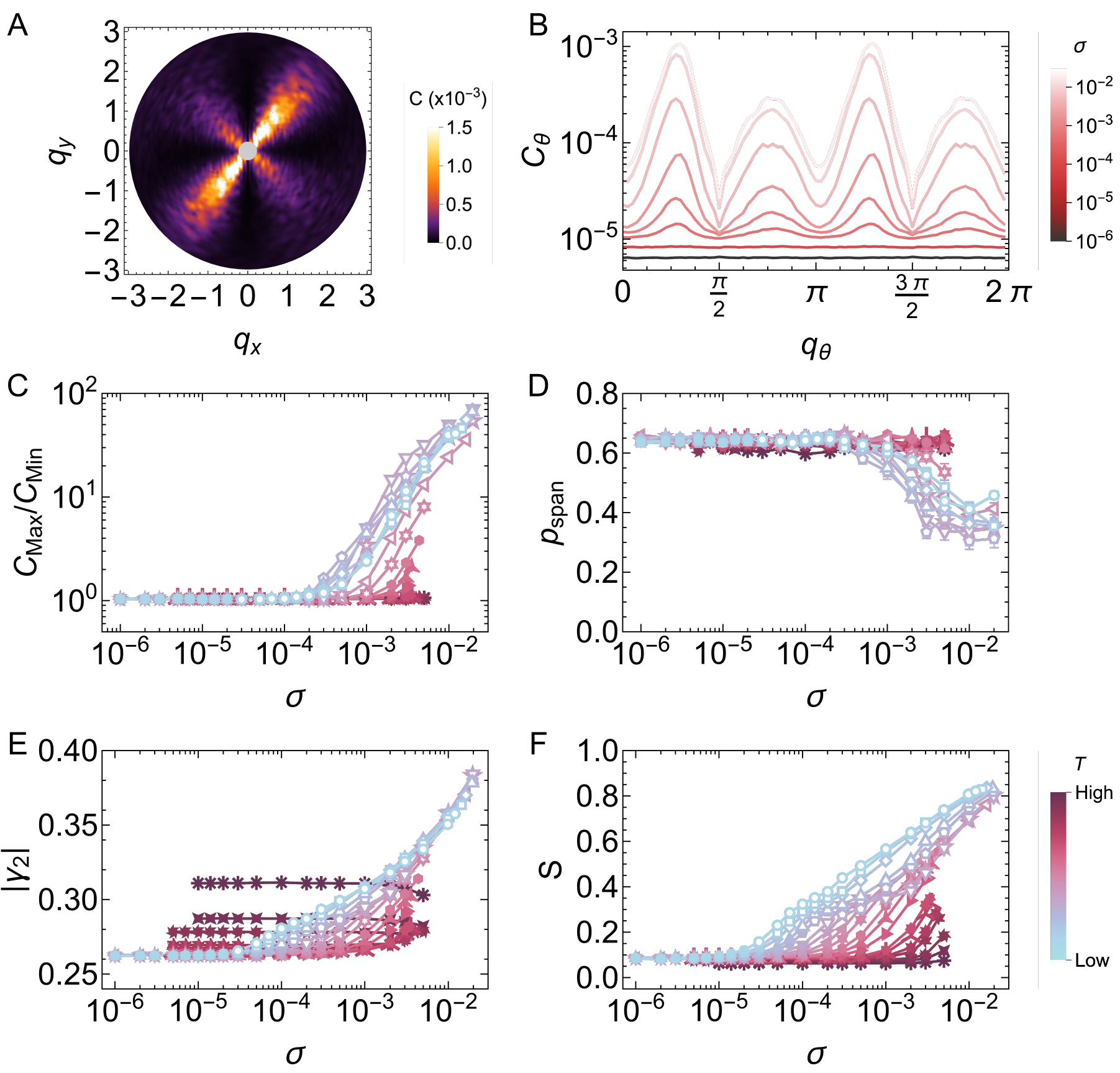}
\caption{
\textbf{Shear jamming is accompanied by distinct structural signatures.}
(A) Shear-stress correlation $C(\vb{q})$ at the lowest $T$ for $p_0=3.85$, $\sigma=0.01$ shows quadrupolar structure. The gray circle indicates $\abs{\vb{q}}<q_{\text{min}}$, with $q_{\text{min}}=2\pi/27$ corresponding to the shortest accessible wavenumber.
(B) $q_{\theta}$-dependence of $C_{\theta}$, the radially averaged $C(\vb{q})$ for different $\sigma$.
(C) Ratio of the maximum to the minimum of $C_{\theta}$ as $\sigma$ and $T$ are varied.
(D) Minimum fraction of high-tension edges required to form a system-spanning network.
(E) Cell elongation quantified by the shape function $\abs{\gamma_2}$.
(F) Nematic order parameter $S$.
Open markers in C--F indicate state points in the discontinuous shear thickening and shear jammed regimes.
All data is for $p_0=3.85$. 
}
\label{fig:sj-corr}
\end{figure}

\emph{Shear stress correlations---}Anisotropy in the shear stress correlations is a key signature of a jammed or shear-jammed state~\cite{Maloney2006, Henkes2009,Wang2020,Nampoothiri2020,Nampoothiri2022,Vinutha2023,Countryman2025}. 
We characterize this via $C(\vb{q})$, the spatial correlation of local shear stresses in Fourier space (see Methods). 
In the shear-jammed regime (see \cref{fig:sj-corr}A for a representative state point), we find $C(\vb{q})$ has the expected anisotropic quadrupolar structure and appears consistent with the formation of a ``pinch-point'' singularity as $q\to 0$, although thermal effects slightly blur the structure compared to the athermal limit.
This observation provides a qualitative indicator that the structure is indeed rigid and contains the system-spanning force chains expected in a (shear-)jammed system~\cite{Vinutha2023,Countryman2025}.

By radially averaging $C(\vb{q})$ to isolate the angular dependence $C_{\theta}(q_{\theta})$, we can track the emergence of these anisotropic correlations as $\sigma$ is varied (see \cref{fig:sj-corr}B).
We note that although our results are symmetric under the transformation $q_{\theta}\to q_{\theta}+\pi$, we do observe an asymmetry between the peak heights at $q_{\theta}\sim \pi/4$ and $q_{\theta}\sim 3\pi/4$, indicating anisotropy in the underlying structure.
To explore how the $q_{\theta}$-dependence of $C_{\theta}$ emerges across temperatures, we examine the ratio $C_{\text{Max}}/C_{\text{Min}}$ between the maximum and minimum of $C_{\theta}$ for a given state point.
The results (\cref{fig:sj-corr}C and \cref{fig:supp-corr-ratios}) demonstrate that the strong anisotropies in $C(\vb{q})$ emerge only in the shear thickening and shear jammed regimes.

\emph{Cell edge tensions---}The tension $\vb{T}_{ij}$ along an edge shared by cells $i$ and $j$ can be calculated directly from the energy functional (see Methods). 
Building on a recent study that qualitatively highlighted that DST is associated with the formation of system-spanning networks of high-tension edges~\cite{Hertaeg2024}, here we quantify these percolating edge networks.
We determine the minimum proportion $p_{\text{span}}$ of the highest-tension edges that must be included for the resulting network to span the system from top to bottom.
The results, shown in \cref{fig:sj-corr}D and \cref{fig:supp-tensions}, indicate that $p_{\text{span}}\approx 0.65$ in the Newtonian and shear thinning regimes, which is consistent with the random bond percolation threshold on a hexagonal lattice~\cite{Sykes1964}. 
Given that the topology of the cell system is expected to be similar to a hexagonal lattice, this suggests that the high-tension edges are randomly distributed.
However, in the DST and shear jammed regimes, $p_{\text{span}}$ notably decreases, indicating spatial correlations in the distribution of high-tension edges so that fewer edges are required to form system-spanning networks.
Unlike the compressive frictional contacts that drive shear jamming in dense suspensions, these system-spanning networks are strictly tensile.

\emph{Cell shape anisotropy---}The approach to shear jamming is also associated with cell-shape changes, which we quantify through the shape function $\gamma_p$~\cite{Armengol2023} (see Methods).
We characterize cell elongation via $\abs{\gamma_2}$, which is constant at small $\sigma$ in the Newtonian and shear-thinning-dominant regimes, 
but substantially increases above the onset of shear thickening (see \cref{fig:sj-corr}E and \cref{fig:supp-cell-shape}).
This is accompanied by a decrease in $\abs{\gamma_6}$ (see \cref{fig:supp-cell-shape}), which quantifies how close the cells are in shape to regular hexagons.
The increase in $\abs{\gamma_2}$ also coincides with the emergence of cell-cell alignment, which we quantify using the nematic order parameter $S$ (see Methods).
\Cref{fig:sj-corr}F and \cref{fig:supp-cell-shape} highlight the high degree of nematic ordering that accompanies cell elongation. 
Note that although low $p_0$ also displays a systematic increase in $\abs{\gamma_2}$ and $S$ at high $\sigma$, it is much less pronounced than at higher $p_0$, indicating that the elongation and alignment are primarily associated with shear thickening and shear jamming.

\section*{Discussion}

Our unified scaling framework (\cref{eq:scaling}) captures the diverse nonlinear rheology of the Voronoi cell model,
showing that distinct flow regimes, one dominated by yielding and shear thinning and the other by shear thickening and shear jamming, are governed by the underlying competition between relaxation and external driving.
This framework relies on a power-law divergence of the structural relaxation time $\tau_{\alpha}\sim(T-T_c)^{-\Delta}$, which strongly contrasts with the typical Arrhenius scaling observed in most disordered or glassy materials.

While the shear thickening regime ($p_0>p_0^*$) shares macroscopic features with dense suspensions~\cite{Ramaswamy2023,Malbranche2022Frontiers,Ramaswamy2025,Barth2026}, their microscopic mechanisms differ: suspensions possess frictional contacts~\cite{Wyart2014,Brown2014,Sharma2026,Bhowmik2026} that are absent in space-filling tissues.
Instead, our ansatz suggests these distinct rheological regimes are governed purely by the competition between external driving and structural relaxation.
Consequently, structural signatures (such as cell elongation, edge tension networks, and stress correlations) emerge from slowing dynamics rather than acting as the microscopic origin of the transition.
Furthermore, our framework captures the shear thinning regime, which is a common physical feature~\cite{Westermeier2016,Gluszek2019,Singh2019} that has been excluded in comparable scaling studies in dense-suspensions~\cite{Malbranche2022Frontiers}.

A further point of comparison is the relative prevalence of shear thinning and shear thickening as $p_0$ is varied.
In dense suspensions, shear thickening is expected to dominate, yet shear thinning is much more readily observed~\cite{Barnes1989,Brady1985,Brown2010}.
This has been attributed to a yield stress masking the underlying thickening regime; tuning the inter-particle interactions to reduce $\sigma_y$ allows materials to shear thicken~\cite{Brown2010,Pednekar2017,Richards2021}, in analogy to increasing $p_0$.
Similarly, while shear thinning has been suggested in living tissues~\cite{Sanematsu2021,kiran2021effect}, experimental evidence for macroscopic shear thickening remains elusive.
Our framework suggests an explanation: many experiments investigating tissue rheology are performed on mature, stable epithelia that naturally reside in the crowded, solid-like regime characterized by low $p_0$~\cite{Park2015,Farhadifar2007}.
This implies that experiments must tune the baseline tissue fluidity to target a shear thickening transition. 
Candidate systems include unjammed monolayers prior to crowding-induced solidification~\cite{chisolm2025transitions}, and actively flocking, fluidized collectives, such as those driven by endocytic reawakening~\cite{malinverno2017EndocyticReawakeningMotility,palamidessi2019UnjammingOvercomesKinetic}.

In the context of epithelial tissues, our framework provides a unified physical mechanism for how monolayers might dynamically regulate their macroscopic mechanics to suit distinct physiological needs.
For tissues dominated by strong cortical tensions, the baseline solid-like state allows them to act as robust physical barriers~\cite{charras2014PhysicalInfluencesExtracellular,guillot2013MechanicsEpithelialTissue}.
In these systems, the observed yielding and subsequent shear thinning could provide a physical route to temporarily fluidize, accommodating the extensive collective rearrangements required during processes like wound healing and morphogenesis~\cite{mongera2018fluid,petridou2021RigidityPercolationUncovers}.
Conversely, for tissues that are naturally more fluid-like, shear thickening and shear jamming could serve as a mechanical fail-safe: when subjected to sudden, large physiological shear stresses, the monolayer can spontaneously solidify to resist catastrophic deformation.
Furthermore, such mechanisms could act as a critical mechanical signal; macroscopic tissue stiffening is known to be a functional prerequisite for initiating developmental programs such as the epithelial-mesenchymal transition \emph{in vivo}~\cite{Barriga2018}.

While the diverse flow regimes observed share striking similarities with dense suspensions, our results show the underlying microscopic mechanisms and scaling frameworks are fundamentally distinct~\cite{Ramaswamy2023}.
This emphasizes that confluent tissues represent a distinct class of disordered matter, and underscores the need for theoretical models that go beyond suspension-based paradigms.
Ultimately, linking these universal scaling predictions to experimental measurements of tissue rheology offers a promising route toward understanding how microscopic cellular properties evolved to regulate macroscopic mechanical function in living systems.

\section*{Methods}
\subsection*{Simulation details}
We perform simulations in the canonical (NVT) ensemble using the open source software package, \textit{cellGPU}~\cite{Sussman2017}.
We set $K_A=1$ and $K_P=1$, and set the unit of length $l=\sqrt{A_0}=1$. 
We perform simulations in a periodic box of side length $L=\sqrt{N}$ to allow all cells to achieve their target area.
Thermostatting is achieved by coupling the system to a Nos\'e-Hoover chain~\cite{Martyna1992,Martyna1996}.
For each simulation, we allow an initially high temperature configuration to equilibrate at a temperature $T$ for $\sim10\tau_{\alpha}$ before turning on the shear. 

We apply shear using a reverse nonequilibrium molecular dynamics (RNEMD) technique in which a positive momentum flux (shear stress) is applied to a narrow horizontal slab of width $l$ at the center of the system ($y=L/2$), while a negative momentum flux is applied to a slab at the bottom ($y=0$)~\cite{Kuang2012,MullerPlathe1997,MullerPlathe1999,Tenney2010,Kuang2010}. 
In regimes where the system is fluid-like, this induces a linear velocity gradient $v_x(y)$ in each half of the system, but with no overall net flow (see schematic, \cref{fig:schematic}).
Here, we apply the velocity shearing and scaling (VSS) RNEMD algorithm proposed by Kuang and Gezelter~\cite{Kuang2012}.
While the momentum exchange is artificial, VSS-RNEMD provides an effective technique for studying stress-controlled rheology of simulated systems while preserving the total linear momentum, total kinetic energy, and Maxwell-Boltzmann velocity distribution. 
Note that these physical constraints set the upper limit on $\sigma$ values we are able to investigate using this approach.

We set the shear updater period to $\Delta t=5 \dd t$, where $\dd t =0.01\tau$ is our simulation time step and $\tau$ is the simulation time unit, and allow the dynamics to run for a further $50\,000\tau$ to ensure the system has reached steady state.
We then record data for at least $60\,000\tau$.
Our results are not sensitive to the choice of $\Delta t$ within a reasonable range.

The viscosity $\eta = \sigma/\dot\gamma$ is determined from the applied shear stress $\sigma$ and the measured flow rate $\dot\gamma=\dv{v_x}{y}$, which for a given sample is determined by performing a linear fit on the velocity profile in each half of the system and averaging the two after flipping the sign in one half.
In solid-like regimes the cells do not flow, and the velocity profiles do not attain a linear velocity gradient in each half of the system.
We classify points as solid, meaning a macroscopic viscosity cannot be determinied, if the fits to the velocity profile in each half of the system do not have the correct sign, which indicates the velocities are distributed around zero.
Note that close to yielding and shear jamming, finite-time measurements at a given state point can fluctuate between flowing and arrested states. 
We report viscosity values exclusively for state points that reliably exhibit finite $\eta$ across all independent configurations.

We set $N=4096$ throughout and we have verified that our results for select data points are consistent with data from system sizes up to $N=32,768$. 
For smaller system sizes, we do observe strong system-size effects, as discussed in the Results.
All state points are averaged over 10 independent configurations.

\subsection*{Classifying the flow regimes}
To classify the different flow regimes we introduce 
\begin{equation}
G = \dv{\log{\dot\gamma}}{\log{\sigma}}.
\end{equation}
In this definition, Newtonian flow has $G=1$, while $G>1$ for shear thinning and $G<1$ for shear thickening.
We determine $G$ by examining successive pairs of $\sigma$ values sampled, using the flow curves in \cref{fig:supp-gamma-sigma}.
We classify $G> 1.1$ as shear thinning, $0.2 < G < 0.8$ as CST and $G \leq 0.2$ as DST, meaning that $0.8\leq G\leq 1.1$ is classified as Newtonian.
These thresholds are comparable to those used in Ref.~\cite{Hertaeg2024}, and are consistent with visual observations of the flow curves.
Note that in the low-$\sigma$ regime, especially close to yielding, $\eta(\sigma)$ can fluctuate between sampled points, leading to some points being classified as shear thickening in an otherwise shear thinning regime. 
In these cases, we classify these points in line with the overall apparent flow regime in the vicinity of the point.

\subsection*{Phase features of the unsheared system}
We characterize the phase of the unsheared system through the orientational and translational order parameters $\Psi_6$ and $\Psi_T$, and their corresponding susceptibilities $\chi_{\alpha} = \expval{\Psi_{\alpha}^2} - \expval{\Psi_{\alpha}}^2$ for $\alpha = 6, T$. 
Here the $\Psi_{\alpha}$ are the values of the order parameters averaged over the entire system, i.e. $\Psi_{\alpha} = \frac{1}{N}\sum_{j=1}^N \Psi_{\alpha,j}$.
The orientational order parameter for cell $j$, $\Psi_{6,j}$, is defined by
\begin{equation}
\Psi_{6,j} = \frac{1}{N_j}\sum_{n=1}^{N_j} \mathrm{e}^{6i\theta_n},
\end{equation}
where $N_j$ is the number of nearest neighbors and $\theta_n$ is the angle between the vector connecting the centers of cells $j$ and $n$ and a fixed arbitrarily chosen axis.

The translational order parameter for a cell $j$ is
\begin{equation}
\Psi_{T,j} = \mathrm{e}^{i\vb{G}\vdot\vb{r}_j},
\end{equation}
where $\vb{G}$ is a reciprocal lattice vector and $\vb{r}_j$ is the cell position. 
To find $\vb{G}$, we calculate the 2D static structure factor $S(\vb{k})$ (see \cref{fig:supp-sf}), and set $\vb{G}$ as the location of the largest peak of $S(\vb{k})$.
Following Ref.~\cite{Han2008}, we calculate $\Psi_T$ for several values of $\vb{G}$ close to the peak of $S(\vb{k})$ and use the value that maximizes $\Psi_T$ at a given state point. 
For state points with translational ordering this will ensure the translational ordering is captured, while for those without translational ordering this will result in the order being investigated along an arbitrary direction.
The susceptibilities $\chi_{\alpha}$ are then calculated across ten independent trajectories, with at least ten frames spaced by intervals of $\sim\tau_{\alpha}$ included from each trajectory.

\subsection*{Characterizing structural signatures}
\emph{Shear stress correlations---}For a given cell, the shear stress $\sigma_i \equiv \sigma_{xy,i}$ is given by $\sigma_i = \frac{1}{A_i} \pdv{E_i}{\gamma}$.
In Fourier space, correlations in the shear stress are $C(\vb{q}) = \expval{\Delta\sigma(\vb{q})\Delta\sigma(-\vb{q})}$, where $\Delta\sigma(\vb{q}) = \sum_{j=1}^N \left(\sigma_{j}-\expval{\sigma}\right)\mathrm{e}^{i\vb{q}\vdot\vb{r}_j}$.
The shearing protocol means structural features in $C(\vb{q})$ are out of phase between the two system halves; we calculate $C(\vb{q})$ in each half before averaging.
To avoid artifacts from cells in the sheared regions, we exclude cells in the manipulated slabs and in a region of width $l$ either side of those slabs, leading to $q_{\text{min}}\approx 2\pi/27$.

\emph{Cell edge tensions---}The tension $\vb{T}_{ij}$ along an edge with unit vector $\vu*{\ell}_{ij}$ shared by cells $i$ and $j$ is 
\begin{equation}
\vb{T}_{ij} = \pdv{E}{\vb*{\ell}_{ij}} = 2 K_P (P_i + P_j - 2 P_0)\vu*{\ell}_{ij}.
\end{equation}

\emph{Cell shape anisotropy---}The shape function $\gamma_p$ is defined for each cell as~\cite{Armengol2023} 
\begin{equation}
\gamma_p = \frac{\sum_v \abs{\vb{r}_v}^p \mathrm{e}^{ip\phi_v}}{\sum_v \abs{\vb{r}_v}^p}.
\end{equation}
The magnitude $0\leq\abs{\gamma_p}\leq 1$ quantifies closeness to a $p$-sided polygon.
The nematic order parameter is $S=\expval{\cos{2(\theta_i - \theta_{N_i})}}$, where $\theta_i = \mathrm{Arg}(\gamma_2)_i/2$ is the cell orientation and $\theta_{N_i}$ is the average orientation of its neighbors.

\section*{Acknowledgements}
We thank Chengling Li for providing the $\tau_\alpha$ data.
We thank Bulbul Chakraborty, Anna Barth, Paarth Gulati, Sean Ridout, and Eric Weeks for discussions.
HSA acknowledges funding from the Tarbutton Postdoctoral Fellowship.
This material is based upon work supported by the National Science Foundation under Grant No. DMR-2143815 and was supported in part by grant NSF PHY-2309135 to the Kavli Institute for Theoretical Physics (KITP).
This research used the Delta advanced computing and data resource which is supported by the National Science Foundation (award OAC 2005572) and the State of Illinois. Delta is a joint effort of the University of Illinois Urbana-Champaign and its National Center for Supercomputing Applications.

\bibliography{voronoi-nonlinear-rheology-sources}


\newpage
\clearpage
\onecolumngrid

\renewcommand{\thefigure}{S\arabic{figure}}
\makeatletter
\renewcommand{\fnum@figure}{FIG. \thefigure}
\makeatother

\renewcommand{\thetable}{S\arabic{table}}
\makeatletter
\renewcommand{\fnum@table}{TABLE \thetable}
\makeatother

\setcounter{figure}{0}
\setcounter{table}{0}
\setcounter{section}{0}
\setcounter{page}{1}


\begin{center}
{\large\bfseries
Supplementary Material: \\ \papertitle}
    
\vspace{0.5cm}
  
Helen S. Ansell, and Daniel M. Sussman\\

\emph{Department of Physics, Emory University, Atlanta, Georgia 30322}
\end{center}

\begin{figure}[h]
\centering
\includegraphics[width=0.9\textwidth]{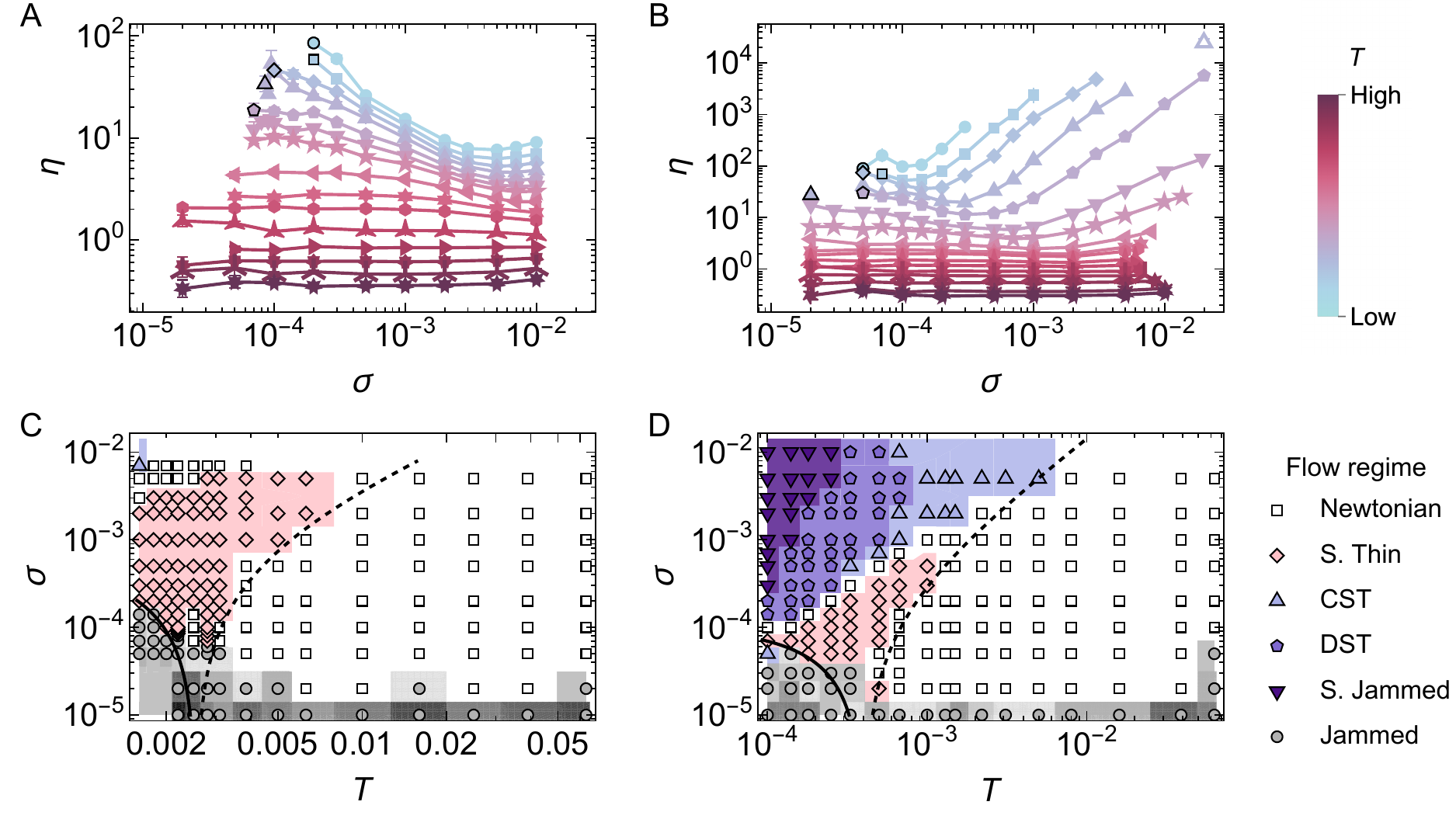}
\caption{Shear-stress-dependence of the viscosity for (A) $p_0=3.80$, and (B) $p_0=3.825$ at different temperatures. The corresponding flow phase diagrams are shown in panels (C) and (D).
Solid black lines in C and D indicate the predicted boundary between solid-like and flowing regimes, while dashed lines indicate the predicted boundary between linear and non-linear flow regimes. 
}
\label{fig:supp-flow-phase}
\end{figure}

\begin{figure}
\centering
\includegraphics[width=\textwidth]{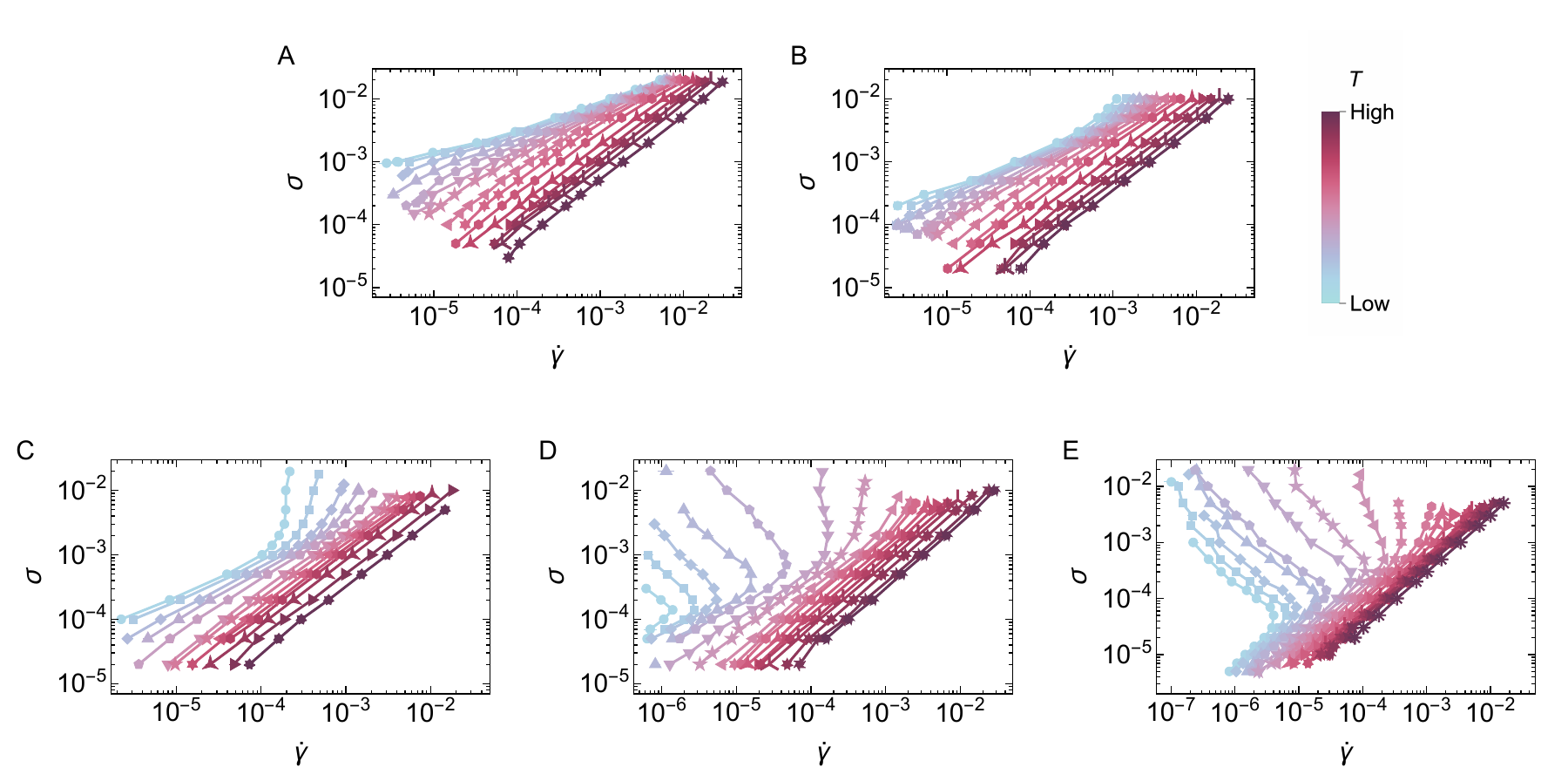}
\caption{Dependence of the measured shear rate $\dot\gamma$ on the applied shear stress $\sigma$ for 
(A) $p_0=3.75$,
(B) $p_0=3.80$,
(C) $p_0=3.815$,
(D) $p_0=3.825$, and
(E) $p_0=3.85$
for different temperature values.
A finite yield stress is characterized by a finite-$\sigma$ intercept of the axes at small $\dot\gamma$. 
Discontinuous shear thickening (DST) is characterized by discontinuities in the flow curves. 
}
\label{fig:supp-gamma-sigma}
\end{figure}

\begin{figure}
\centering
\includegraphics[width=0.9\textwidth]{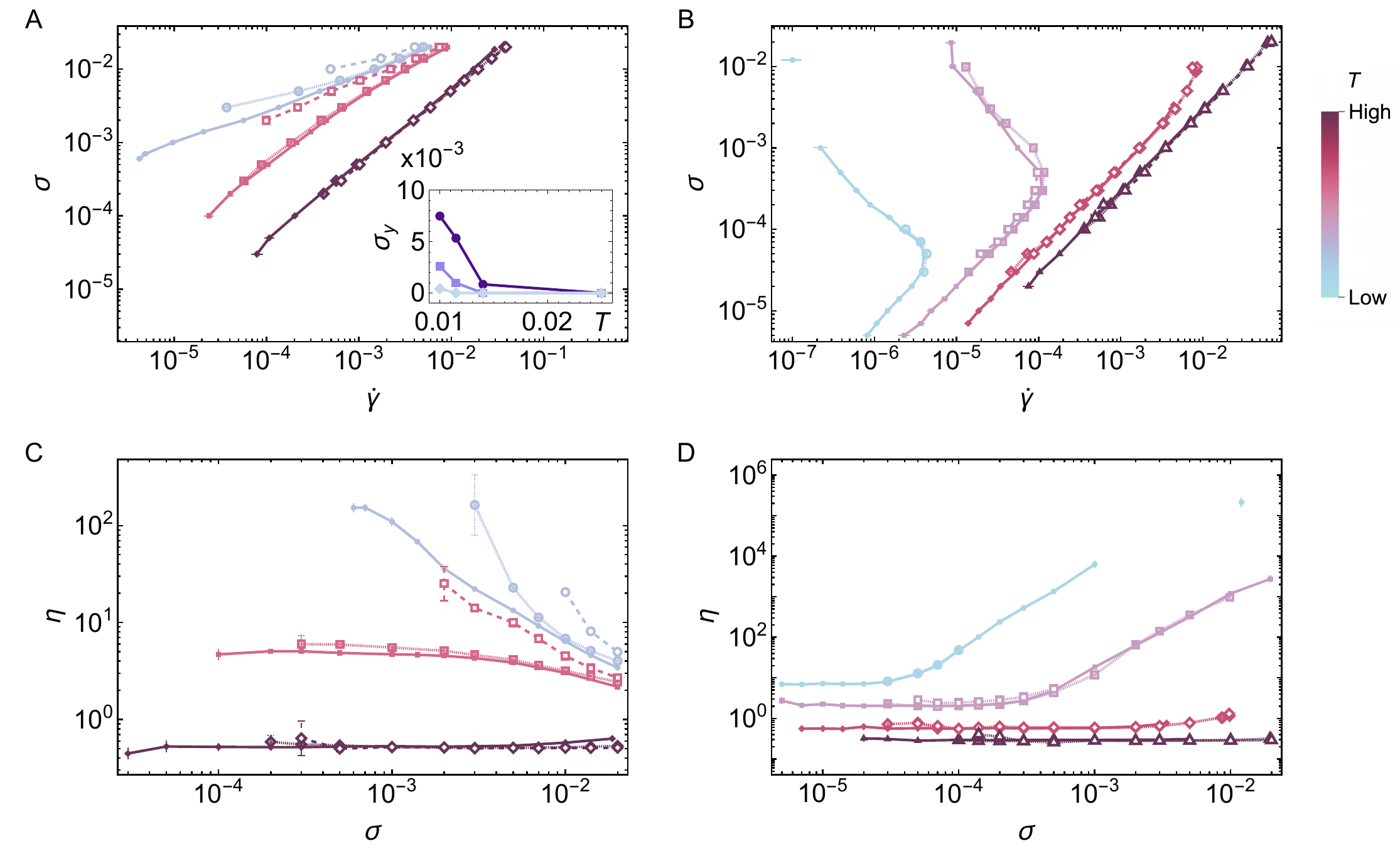}
\caption{Effect of system size on the observed flow behavior.
(A--B) Relation between applied shear stress $\sigma$ and flow rate $\dot \gamma$ for select temperatures for (A) $p_0=3.75$ and (B) $p_0=3.85$.
Data points are shown for $N=128$ (open markers and short dashed lines), $N=512$ (lightly shaded markers and long dashed lines) and $N=4096$ (solid markers and solid lines).
(inset) Measured yield stress $\sigma_y$, determined by fitting a Herschel-Bulkley model, $\sigma = \sigma_y + k \dot{\gamma}^n$, to data points in the main panel. 
Data points are for $N=128$ (circle markers), $N=512$ (square markers) and $N=4096$ (diamond markers).
In the thermodynamic limit, we anticipate the finite yield stress remains at the lowest temperatures.
(C--D) Corresponding plots of measured shear viscosity $\eta$  as $\sigma$ is varied.
For $p_0=3.85$, the system does not flow at the lowest $T$ values for $N=128$.
Plot colors and plot marker shapes are the same for a given $T$ for each $N$. 
}
\label{fig:supp-fss}
\end{figure}

\begin{figure}
\centering
\includegraphics[width=0.9\textwidth]{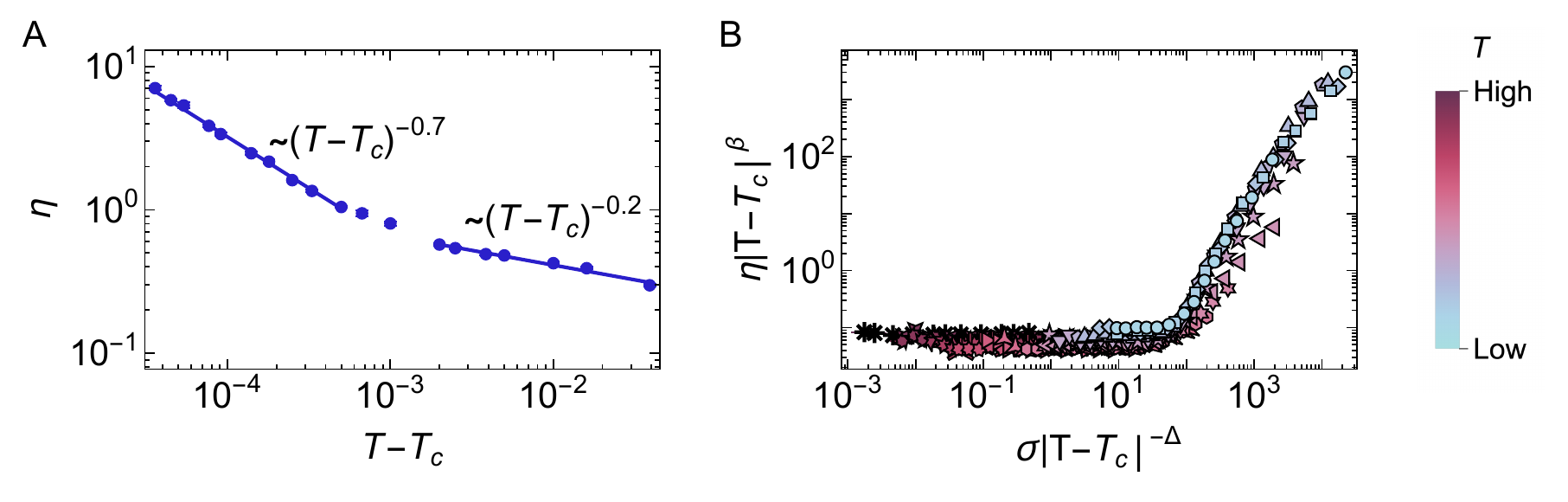}
\caption{
(A) Temperature-dependence of the viscosity $\eta$ in the low-$\sigma$ limit, for $p_0=3.85$ with $T_c=0$. The data exhibits two different power-law regimes.
(B) Best scaling collapse of the $p_0=3.85$ data using \cref{eq:scaling} but with the exponent $\delta=0$. This gives $\beta=0.42$.
}
\label{fig:supp-p385-coll-no-q}
\end{figure}

\begin{figure}
\centering
\includegraphics[width=\textwidth]{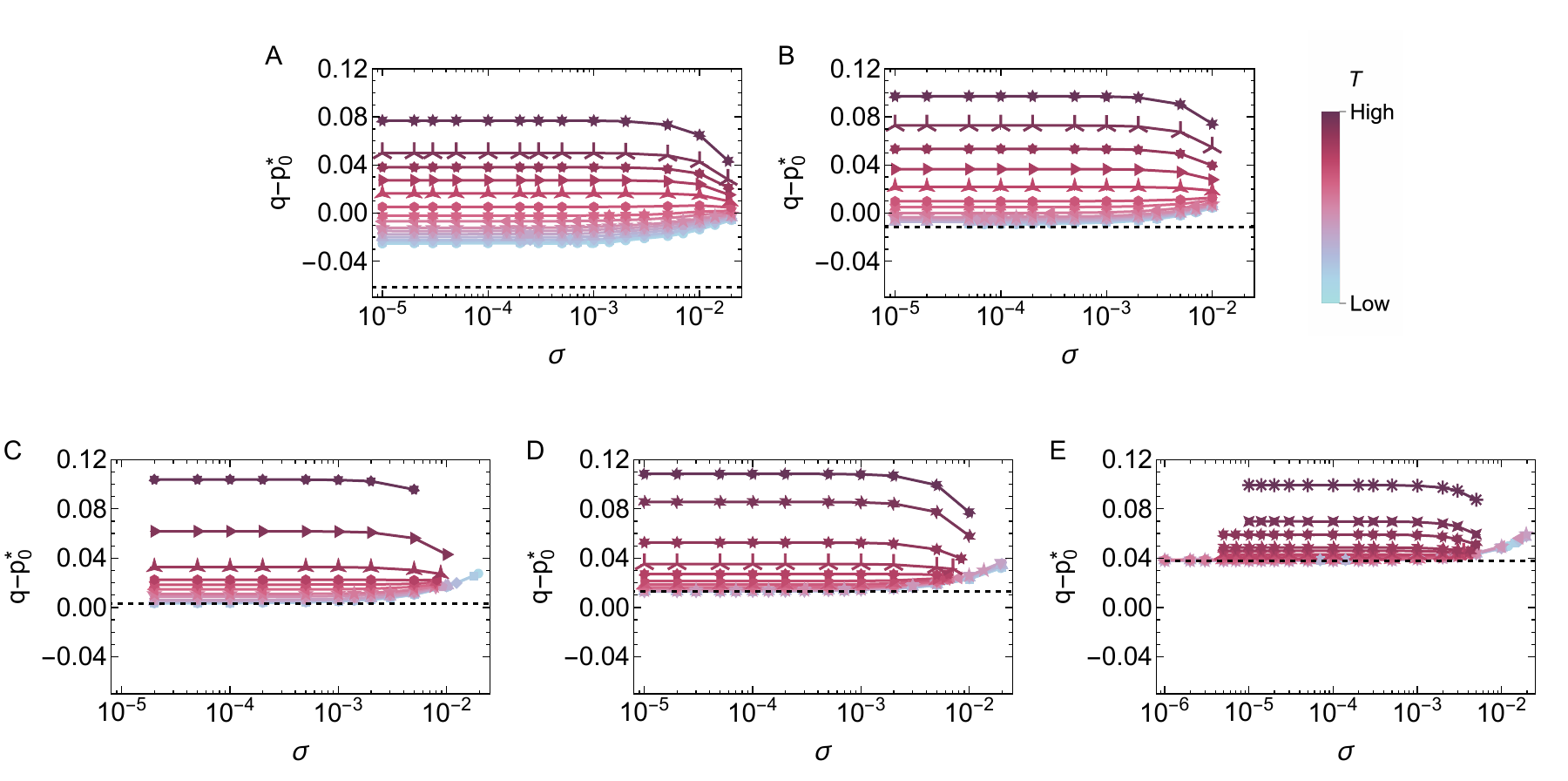}
\caption{Shear-stress-dependence of $q-p_0^*$, where $q$ is the mean measured cell shape index and $p_0^*\approx 3.81$ for 
(A) $p_0=3.75$,
(B) $p_0=3.80$,
(C) $p_0=3.815$,
(D) $p_0=3.825$, and
(E) $p_0=3.85$.
Line colors correspond to different temperatures, as indicated by the color bar.
Dashed black lines indicate $p_0-p_0^*$.
}
\label{fig:supp-q}
\end{figure}

\begin{figure}
\centering
\includegraphics[width=\textwidth]{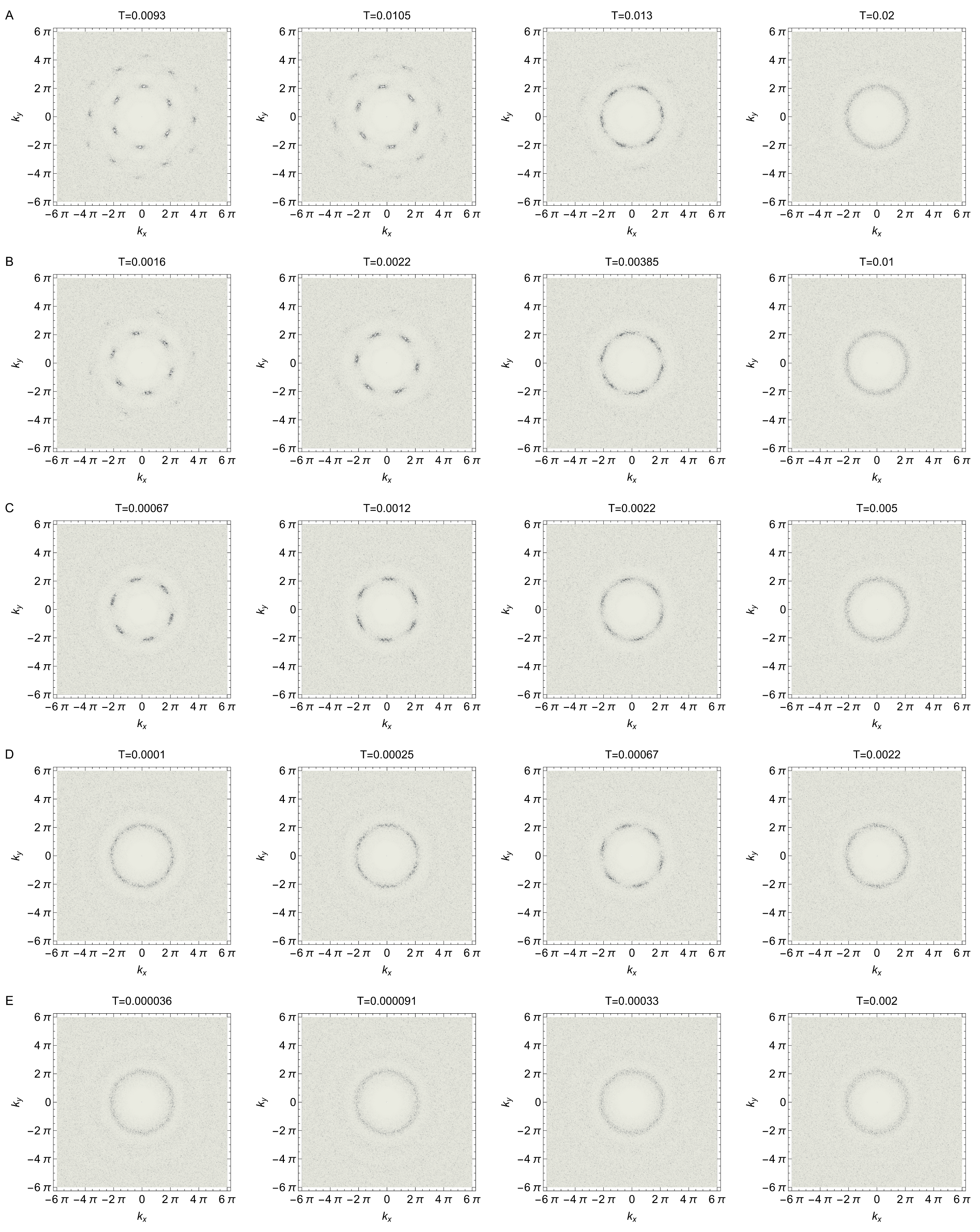}
\caption{Plots of the static structure factor $S(\vb{k})$ in the unsheared system for
(A) $p_0=3.75$,
(B) $p_0=3.80$,
(C) $p_0=3.815$,
(D) $p_0=3.825$, and
(E) $p_0=3.85$.
Panels show a range of temperatures for each $p_0$, as indicated.
}
\label{fig:supp-sf}
\end{figure}

\begin{figure}
\centering
\includegraphics[width=0.9\textwidth]{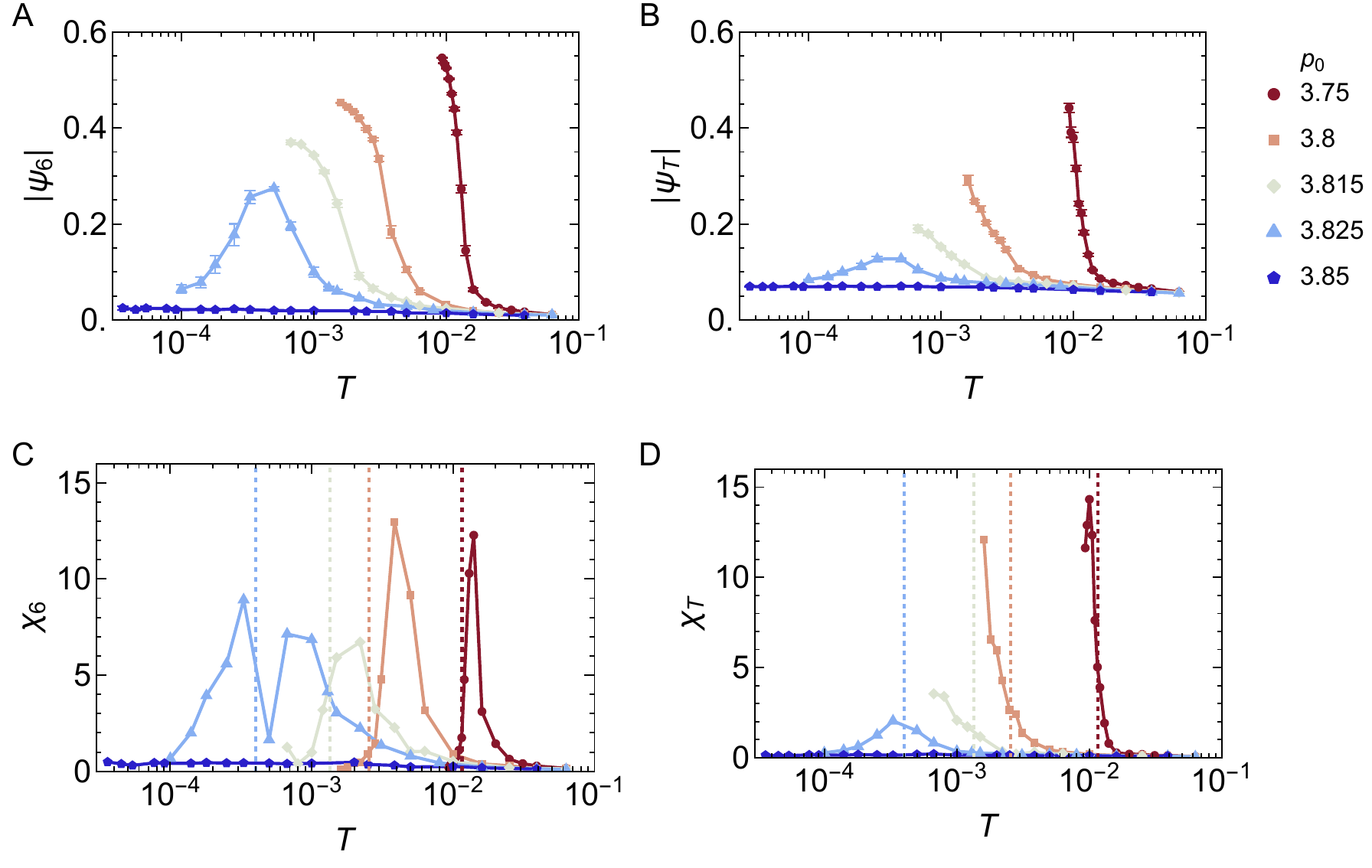}
\caption{Dependence of 
(A) the magnitude of the bond orientational order parameter $\abs{\Psi_6}$, 
(B) the magnitude of the translational order parameter $\abs{\Psi_T}$, 
(C) the orientational susceptibility $\chi_6$, and
(D) the translational susceptibility $\chi_T$ 
on temperature $T$ for each $p_0$ value in the unsheared system.
Error bars in A and B indicate the standard error across ten independent trajectories.
Within each trajectory, at least ten snapshots separated by $\sim\tau_{\alpha}$ are used to calculate $\abs{\Psi_{\alpha}}$.
Dashed vertical lines in C and D indicate the $T_c$ values for each $p_0$.
We use the locations of the peaks of $\chi_6$ and $\chi_T$, which we denote $T_6$ and $T_T$ to classify the equilibrium phase of the system. 
If $T>T_6$, the system is fluid-like, $T_T<T<T_6$ indicates the hexatic phase while $T<T_T$ in the solid-like regime.
We find that the $T_c$ values attained via the system rheology sit close to the corresponding $T_T$ values, indicating that $T_c$ coincides with the emergence of macroscopic solid-like behavior.
For $p_0=3.85$, the value $T_c=0$ is consistent with the lack of structural ordering across the full range of $T$ studied.
}
\label{fig:supp-order-param}
\end{figure}

\begin{figure}
\centering
\includegraphics[width=\textwidth]{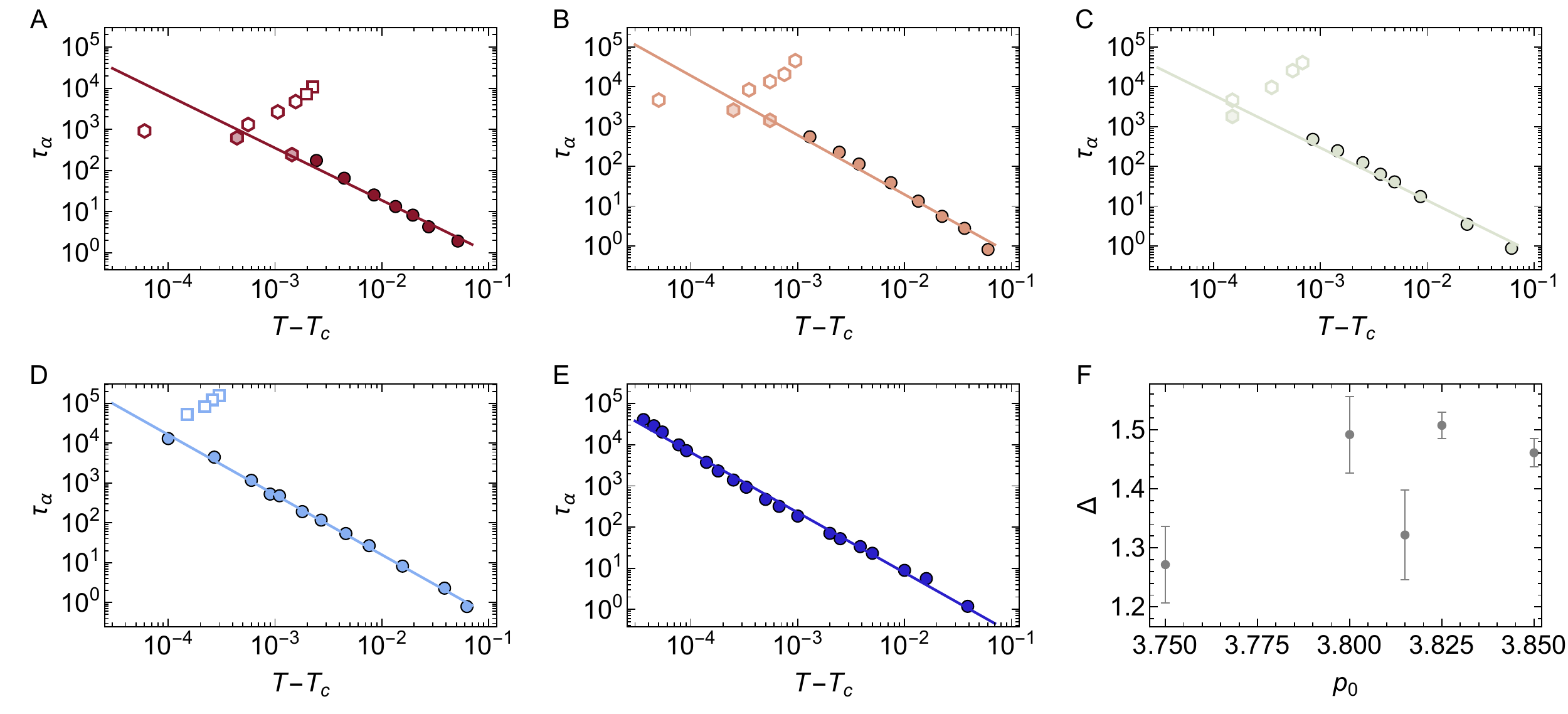}
\caption{Scaling of the $\alpha$-relaxation time $\tau_{\alpha}$ with $T-T_c$ for
(A) $p_0=3.75$, (B) $p_0=3.80$, (C) $p_0=3.815$, (D) $p_0=3.825$ and (E) $p_0=3.85$ in the \emph{unsheared} system.
In each case, points identified as being in the fluid phase are shown as circular markers, those in the hexatic phase are shown with hexagonal markers and those exhibiting translational ordering are shown as square markers (see \cref{fig:supp-order-param}).
Solid markers have $T>T_c$ while open markers have $T<T_c$. 
Plotted lines indicate power law fits of the form $\tau_{\alpha}\sim (T-T_c)^{-\Delta}$ for the $T>T_c$ regime.
(F) Values of the $\Delta$ exponent for each $p_0$. The average value is $\Delta=1.4\pm 0.1$.
}
\label{fig:supp-tau-alpha}
\end{figure}

\begin{figure}
\centering
\includegraphics[width=\textwidth]{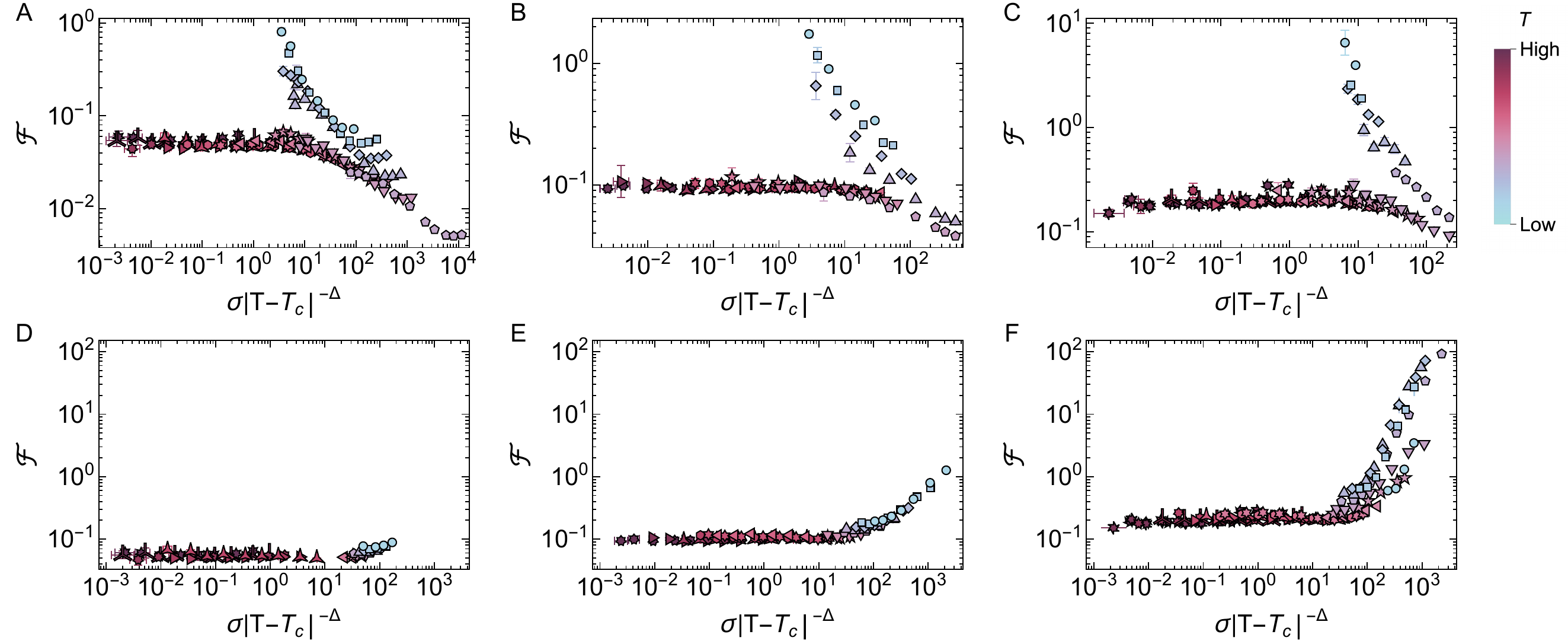}
\caption{Scaling collapse at intermediate $p_0$ values using the scaling ansatz in \cref{eq:scaling}.
(A--C) collapse of the shear thinning and Newtonian regimes for 
(A) $p_0=3.80$, 
(B) $p_0=3.815$, and 
(C) $p_0=3.825$.
Deviations from the low-$T$ branch at larger values of the scaling variable are from points classified as Newtonian that sit at the crossover between the shear thinning and shear thickening regimes.
(D--F) Corresponding collapses of the shear thickening and Newtonian regimes using the same values of the exponents $\beta$, $\Delta$ and $\delta$ for a given $p_0$, but with different (lower) $T_c$ values.
Note that although we observe shear thickening only for a single point in the $p_0=3.8$ phase diagram, the Newtonian and shear thickening data can still be collapsed onto a single curve.
As with higher $p_0$ values, this is achieved by using the same scaling form but with a lower critical temperature ($T_c=0.0006$, compared to $T_c=0.0026$ for shear thinning).
}
\label{fig:supp-scaling}
\end{figure}

\begin{figure*}[h]
\centering
\includegraphics[width=0.9\textwidth]{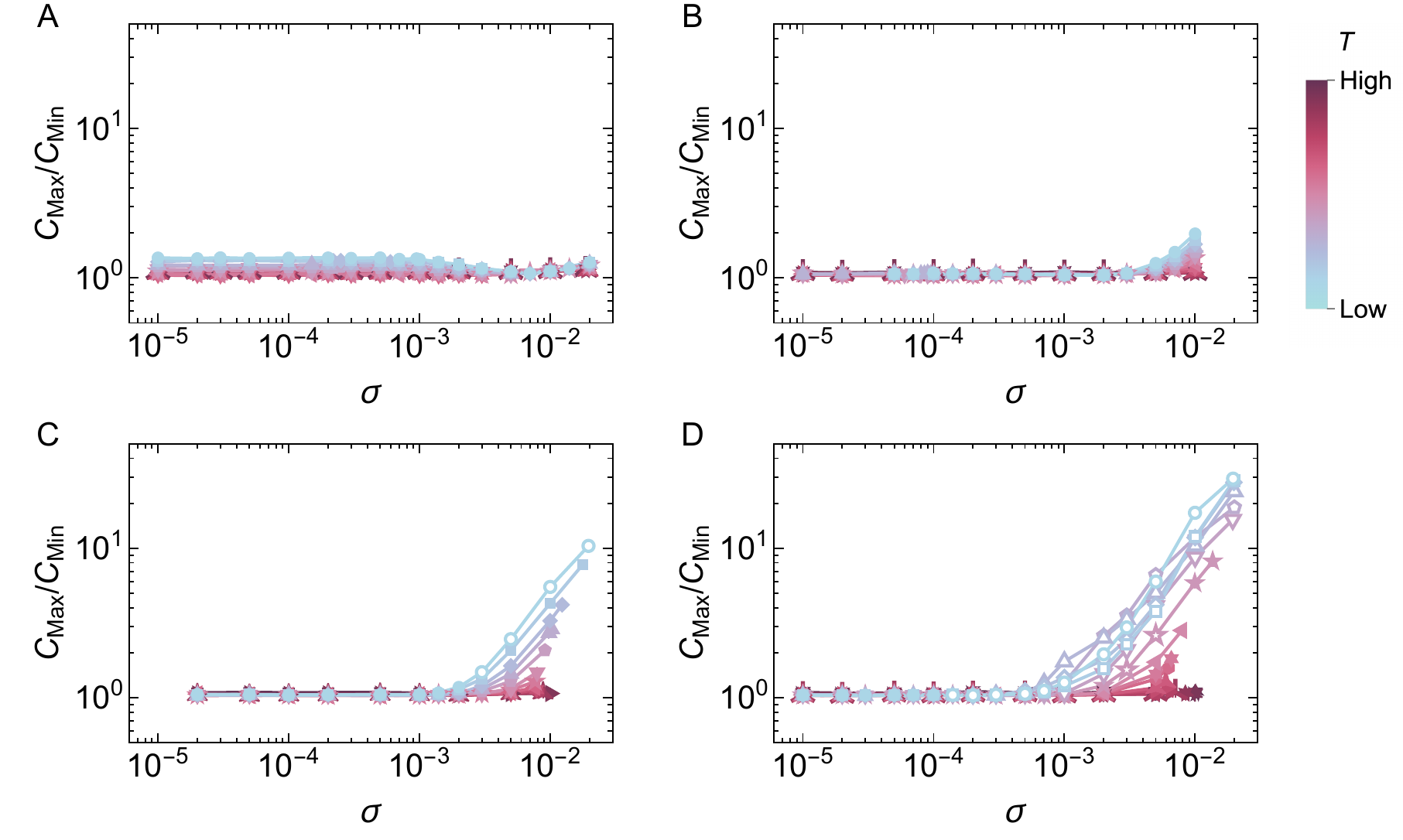}
\caption{Shear-stress-dependence of the ratio of the maximum and minimum of the shear stress correlations $C_{\theta}(q_{\theta})$ for 
(A) $p_0=3.75$,
(B) $p_0=3.80$,
(C) $p_0=3.815$, and
(D) $p_0=3.825$.
Data for $p_0=3.85$ is shown in \cref{fig:sj-corr}C of the main text.
Open markers indicate points in the discontinuous shear thickening and shear jammed regimes.
}
\label{fig:supp-corr-ratios}
\end{figure*}

\begin{figure}
\centering
\includegraphics[width=0.9\textwidth]{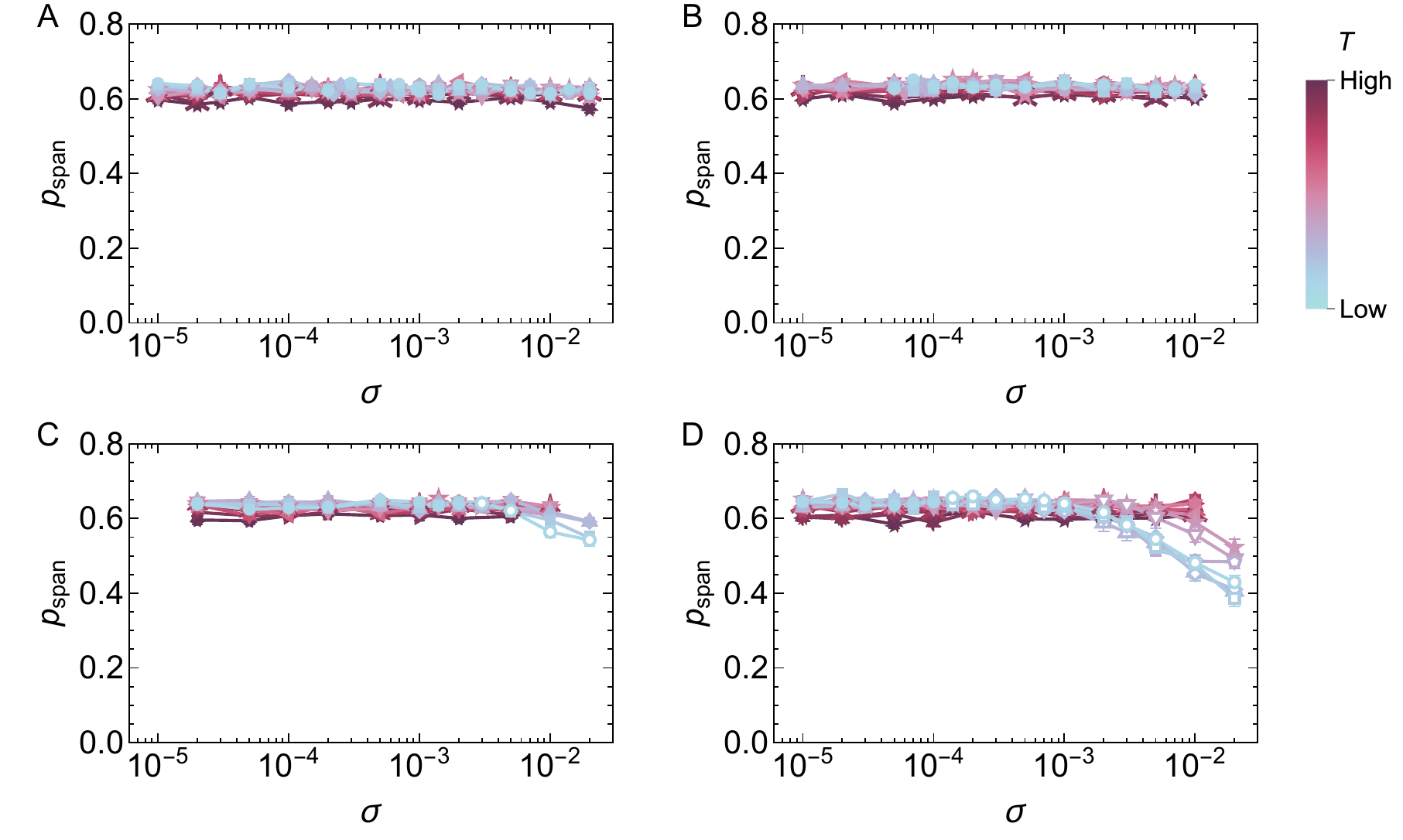}
\caption{Proportion of high-tension edges required for the resulting edge network to span each half of the system for 
(A) $p_0=3.75$, 
(B) $p_0=3.80$, 
(C) $p_0=3.815$ and 
(D) $p_0=3.825$.
Data for $p_0=3.85$ is shown in \cref{fig:sj-corr}D of the main text.
Open markers indicate points in the discontinuous shear thickening and shear jammed regimes.
}
\label{fig:supp-tensions}
\end{figure}

\begin{figure*}[h]
\centering
\includegraphics[width=\textwidth]{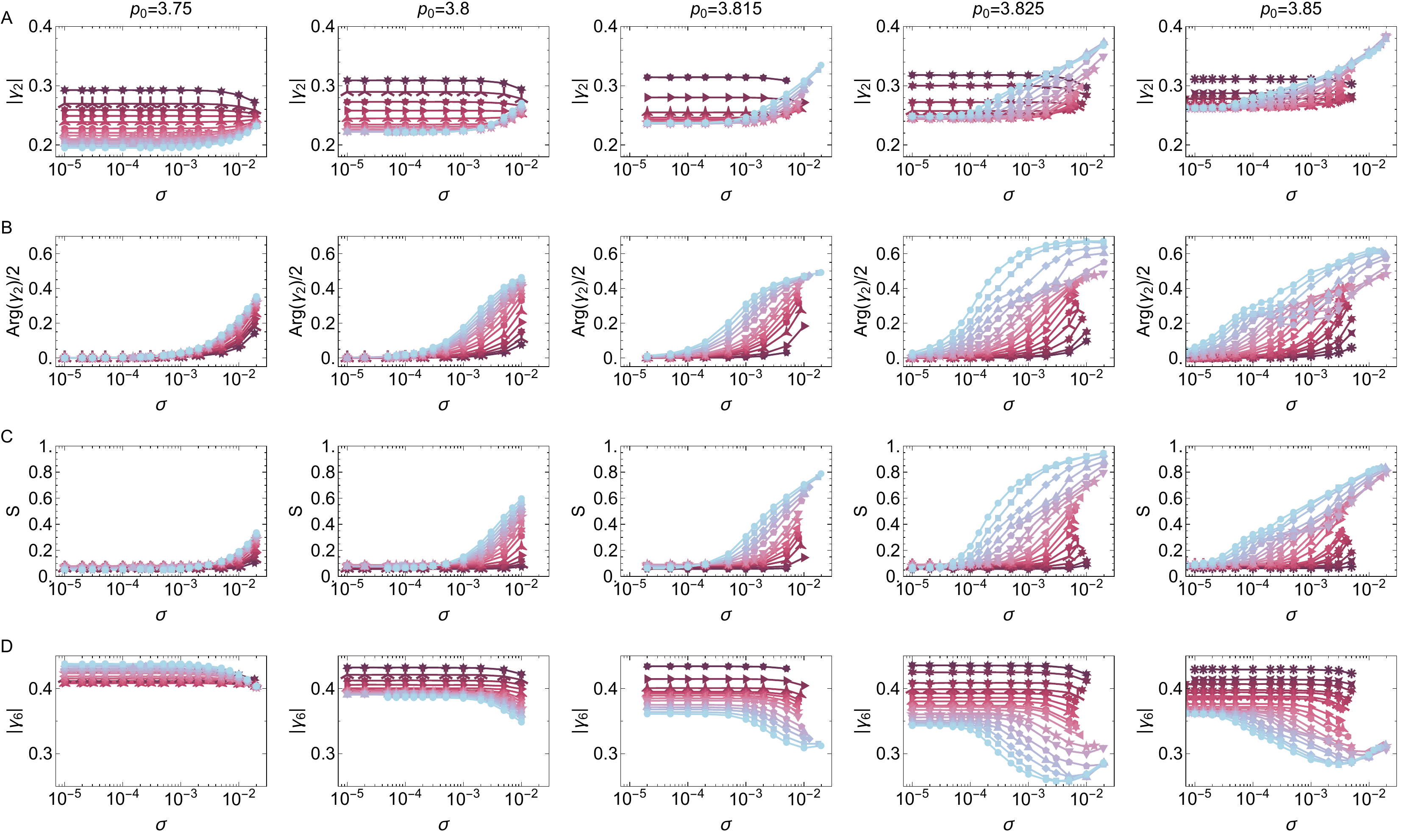}
\caption{Cell-shape analysis across $p_0$ values for different imposed shear stress $\sigma$ and temperature.
(A) Magnitude of the $p=2$ shape function $\abs{\gamma_2}$, which quantifies cell elongation.
(B) Average nematic orientation of cells, with the sign in one half of the system flipped.
(C) Average nematic order parameter $S$.
(D) Magnitude of $\abs{\gamma_6}$, which quantifies how close cells are to being regular hexagons.
Colors in all panels indicate temperature, ranging from low (blue) to high (red). 
}
\label{fig:supp-cell-shape}
\end{figure*}

\begin{figure}[h]
\centering
\includegraphics[width=0.8\textwidth]{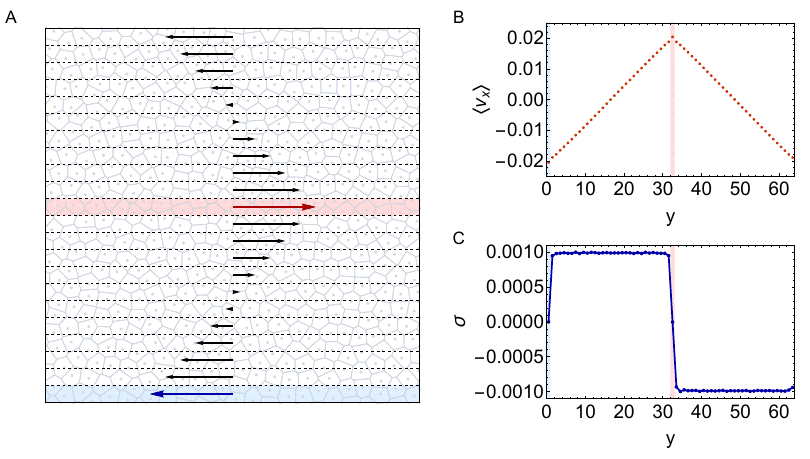}
\caption{
(A) Schematic of the numerical shearing protocol on a snapshot with $N=512$ cells. 
The system is divided into slabs one typical cell size thick, and a flow is induced by manipulating the $x$-direction momenta of cells in the bottom and central slabs (highlighted in blue and red, respectively) to induce a flow in the system while preserving total linear momentum and kinetic energy. Arrows indicate the direction of the induced net flow in each slab. 
(B) Example of a typical velocity profile induced in the system, showing the mean $x$-component of the velocities of cells in each slab for different slabs. The shear rate $\dot\gamma\equiv \pdv{\expval{v_x}}{y}$ is calculated by averaging the gradients of the best fit lines shown in each half of the system, accounting for the sign change.
(C) Example of the resulting shear stress profile induced within the system, measuring the time-averaged value in each slab.
The profiles in B and C are for $N=4096$, $p_0=3.85$, $T=0.001$, $\sigma=0.001$.
}
\label{fig:schematic}
\end{figure}

\end{document}